\documentclass[a4paper,onecolumn,11pt,accepted=2024-06-27]{quantumarticle}
\pdfoutput=1
\usepackage[utf8]{inputenc}
\usepackage[T1]{fontenc}

\usepackage{cite}
\usepackage{xcolor}
\usepackage{braket}
\usepackage{graphicx}
\usepackage{varwidth}
\usepackage{soul}
\usepackage{amssymb,amsmath}
\usepackage{makecell}

\newcommand{\nn}{\nonumber\\}
\DeclareMathOperator{\tr}{tr}
\usepackage{anyfontsize}
\usepackage{dsfont}
\usepackage{tikz}
\usepackage{sidecap}
\usepackage{subfig}
\usepackage{bm}
\sidecaptionvpos{figure}{t}
\usepackage[english]{babel}
\usepackage{enumitem}  

\usepackage{setspace}
\usepackage[permil]{overpic}
\usepackage{mathtools}
\usepackage{wasysym}
\usepackage{placeins}
\usepackage{bm}
\newcommand{\be}{\begin{equation}}
\newcommand{\ee}{\end{equation}}

\definecolor{dur}{cmyk}{0,1,0.6,0.3}
\usepackage{amsfonts}
\usepackage{wrapfig}
\usepackage{braket}
\usepackage{hyperref}
\usepackage{amsthm,verbatim,cancel}
\usepackage{caption}
\newcommand{\bea}{\begin{eqnarray}}
\newcommand{\eea}{\end{eqnarray}}

\DeclareMathOperator{\cluster}{|1\rangle}
\DeclareMathOperator{\sterclu}{\langle 1|}

\newtheorem{Theorem}{Theorem}

\newtheorem{Lemma}{Lemma} 
\newtheorem{Def}{Definition}

\usepackage{ulem}
\newcommand{\bartek}{}

\newcommand{\bart}{}

\begin{document}

\title{The Gauge Theory of Measurement-Based Quantum Computation }
\author[a,b]{Gabriel Wong,} 
\author[c,d]{Robert Raussendorf}
\author[e]{Bart{\l}omiej Czech}

\affiliation[a]{Center for Mathematical Sciences and Applications, Harvard University, Cambridge, MA 02138, USA}
\affiliation[b]{Mathematical Institute, University of Oxford Radcliffe Observatory Quarter, Woodstock Road, Oxford
OX2 6GG, United Kingdom}
\affiliation[c]{Institut f{\"u}r Theoretische Physik, Leibniz Universit{\"a}t Hannover, Appelstra{\ss}e 2, 30167 Hannover, Germany}
\affiliation[d]{Stewart Blusson Quantum Matter Institute, University of British Columbia, Vancouver, Canada}
\affiliation[e]{Institute for Advanced Study, Tsinghua University, Beijing 100084, China}

\abstract{\noindent Measurement-Based Quantum Computation (MBQC) is a model of quantum computation, which uses local measurements instead of unitary gates. Here we explain that the MBQC procedure has a fundamental basis in an underlying gauge theory. This perspective provides a theoretical foundation for global aspects of MBQC. The gauge \bart{transformations reflect} the freedom of formulating the same MBQC computation in different local reference frames. The main identifications between MBQC and gauge theory concepts are: (i) the computational output of MBQC is a holonomy of the gauge field, (ii) the adaptation of measurement basis that remedies the inherent randomness of quantum measurements is effected by gauge transformations. The gauge theory of MBQC also plays a role in characterizing the entanglement structure of symmetry-protected topologically (SPT) ordered states, which are resources for MBQC. Our framework situates MBQC in a broader context of condensed matter and high energy theory. 
 }
 \maketitle

\section{Introduction}\label{Intro}

\begin{figure}
    \centering
    \includegraphics[width=7cm]{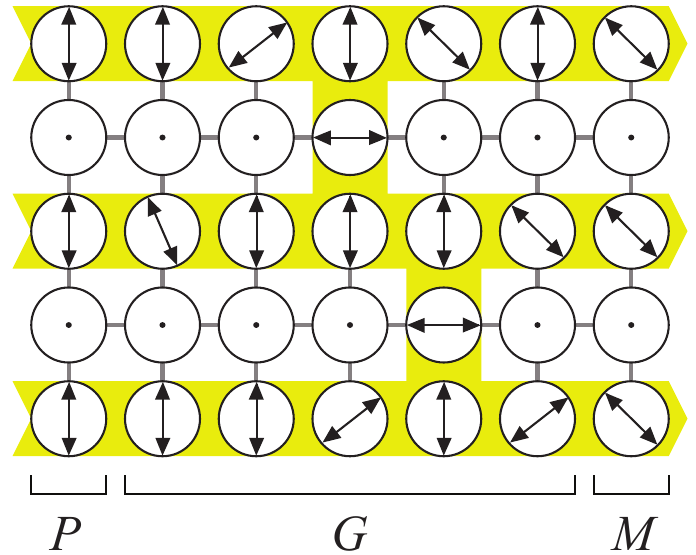}
    \caption{Graphical summary of MBQC with 2D cluster states. The circles represent the individual qubits of the cluster state, and the grey lines define the corresponding entanglement structure. The symbol $\odot$ represents a measurement in the $Z$-basis, and the double-headed arrows are measurements in the eigenbasis of $\cos \alpha X + \sin \alpha Y$, for various angles $\alpha$. The $Z$-measurements have the effect of removing the measured qubits from the cluster, leaving the remainder in a network-like structure. In a circuit model interpretation, the `horizontal' direction represents circuit time (from left to right), and the `vertical' direction labels the logical qubits in the simulated quantum register. The horizontal structures in yellow underlay represent sequences of one-qubit rotations, and the vertical bridges in yellow underlay mediate interaction between logical qubits.}
    \label{fig:scheme}
\end{figure}

Measurement-based quantum computation (MBQC) \cite{RB01} is a scheme of universal quantum computation driven solely by measurements; no unitary evolution takes place in the computational process. The measurements are local, and applied to an entangled initial state---the resource state. The resource perspective (see e.g. \cite{VdN,VdN2}, note \cite{TooE}, however) is suggested by the observation that once all the local measurements have been performed, the post-measurement state is a tensor-product state with no entanglement left. The initial entanglement has been converted into computational output. See Figure~\ref{fig:scheme} for illustration.

The power of MBQC depends on the resource state used. Several types of resource states, such as cluster states \cite{RB01a} and AKLT states \cite{AKLT,AKLT2} on various two-dimensional lattices, give rise to computational universality \cite{RB01,AKLTmbqc_M,AKLTmbqc_W,Darmawan, AKLTmbqc_spin2}. On the other hand, some highly entangled states, such as Kitaev surface code states, have extremely limited computational power for MBQC \cite{Kit}. As a general result, it is known that universal resource state are extremely rare in Hilbert space, for the seemingly paradoxical reason that most states are too entangled to be computationally useful \cite{TooE}. 

The above picture is affected by symmetry. Namely, in certain symmetry protected
phases of matter \cite{GW, Wen1, Wen2, Schuch}, every ground state has the same computational power \cite{Darmawan, SPT0,SPT1,SPT2,SPT3,SPT4,SPT5,SPT5b,SPT6,SPT7}. This phenomenon has been termed `computational phases of quantum matter' \cite{SPT0,SPT1}. Some of these computational phases have universal computational power \cite{SPT5,SPT5b,SPT6,SPT7}. Yet, a classification of universal resource states has to date remained elusive.

Investigations of MBQC have traditionally adopted a local perspective. In this view, a central lemma is that quantum gates can be simulated by applying local measurements to an entangled state. Consequently, every circuit model computation---that is, a series of gates---can be converted into a sequence of local measurements on an appropriate resource state. From this vantage point, MBQC functions as a modular protocol for translating circuit model computations into measurements: gate by gate, measurement by measurement. This locally focused perspective is rightful and important; indeed, this is how the universality of MBQC as a computational model is typically proved. It does not, however, tell the whole story of MBQC.

The present paper lays out a complementary, global perspective. Readers familiar with MBQC will readily identify one aspect of it, which is tellingly non-local: the computational output is a specific combination of measurement outcomes, which were registered over the course of the computation. In this way, the computational power of MBQC relies on correlations between different measurement outcomes, which are extracted over many distinct and generically distant locales in the resource state. This fact is neither here nor there in a local view of MBQC, but it is an essential aspect of the computational scheme! Our quest for a global retelling of the MBQC story was initially motivated by this observation.

One topic in the undergraduate physics curriculum showcases a similar interplay between the local and the global: gauge theory. In the local perspective, defining a gauge theory starts with a vector potential $A_\mu(x)$, which is a local but unphysical quantity. It is unphysical because it is subject to gauge transformations, i.e. local symmetries of the theory. As the local symmetry acts independently at every location, it strips $A_\mu(x)$ of independent physical meaning because it can always locally reset it to an arbitrary different value. But this does not make $A_\mu(x)$ redundant. As famously explained by Aharonov and Bohm, the global object \bartek{${\rm P}\!\exp \oint dx^\mu A_\mu(x)$}---a Wilson loop---measures an unambiguously defined, physical flux. Wilson loops are nonlocal, but they comprise the genuine degrees of freedom of a gauge theory.  These facts show an uncanny resemblance to MBQC. Our task in this paper is to explain that this resemblance is not skin-deep or accidental. We argue that MBQC in all its aspects---global and local---is in fact naturally described by the language of gauge theory. 

There is another reason why gauge theory could be suspected to play a role in MBQC. On one-dimensional cluster states, MBQC can be understood as a sequence of quantum half-teleportations \cite{CLN}. However, as explained in Reference~\cite{withlenny} (see also \cite{mielczarek}), the effect of a quantum teleportation \bartek{can always be mimicked by a background gauge field, which couples to the degree of freedom being teleported. This strongly suggests that it should be possible to conceptualize MBQC as a protocol in gauge theory.}

To make the translation between MBQC and more familiar gauge theories transparent, we highlight the main entries of the proposed dictionary:  
\begin{enumerate}
\item In MBQC, one progressively adjusts measurement bases depending on outcomes of prior measurements. These adjustments represent \bartek{a special gauge transformation, which locally trivializes the connection}; see Section~\ref{TempO}.
\item The computational output of MBQC is a flux analogous to $\oint dx^\mu A_\mu(x)$; see Section~\ref{hf}.
\item Programming MBQC---the choice of quantum algorithm---amounts to choosing a basis, in which the flux is measured; see Section~\ref{sec:quantumgauge}.
\end{enumerate}

The gauge theory perspective is directly connected to symmetry-protected topological (SPT) order; see \cite{Levin2012} for a discussion of how gauge theory characterizes SPT. At the same time, many states in the computational phase of quantum matter, including the one-dimensional cluster state, are SPT-ordered \cite{SPT0, SPT1, SPT2}. In retrospect, the relevance of gauge theory to SPT and the relevance of SPT to MBQC could have motivated a gauge theoretic study of MBQC earlier. In this paper, we uncover a direct relation between gauge theory and MBQC, thereby closing the triangle MBQC-SPT-gauge theory. \bart{SPT order implies that the quantum state is preserved by certain operators, which in the presence of boundaries generate multiplets of edge modes; see e.g. \cite{Yoshida}. In our retelling, those operators generate gauge transformations.}

A final point worth emphasizing is that the present paper is limited to MBQC on a one-dimensional cluster state. This setting allows MBQC to simulate arbitrary $SU(2)$ gates but not larger unitaries, and falls short of universality. The universal, two-dimensional case will be treated in a future publication.

\paragraph{For whom the paper is written, and why} A target audience for this paper is quantum information theorists and condensed matter theorists, who are interested in unifying the various guises of MBQC under a common formalism. MBQC exemplifies many ideas that are of current relevance in high-energy and condensed matter physics, such as holographic duality \cite{ehm, daniel, cohomology} and bulk-boundary correspondence \cite{SPT1, Laughlin, Fuchs}, the emergence of temporal order \cite{BTO,SRTO}, and topological order in 3D \cite{FTQC1,FTQC2,FTQC3,FTQC4}---in the setting of quantum information processing. Recently, the subject of symmetry-protected topological order has contributed to the endeavor of exposing the fundamental structures of MBQC, and here we identify a further ingredient: gauge theory. We have kept the discussion of MBQC self-contained, so physicists outside the quantum information and condensed matter communities should find it easy to digest.

We use the one-dimensional cluster state as a main stage of presentation. This considerably simplifies the discussion while permitting to lay out the general tenet. It shall be noted, however, that universal measurement-based quantum computation requires cluster states in dimension two or higher.

\paragraph{Organization} Section~\ref{sec:review} is a pedagogical review of background material, including the one-dimensional cluster state and the MBQC protocol.
Section \ref{section:prep} sets the stage for the formulation of the MBQC gauge theory by assigning group labels for the MBQC measurement basis using the MPS symmetries of the cluster state.  In Section \ref{sec:mbqcring}, we formulate MBQC on a ring-like cluster state, which manifests most cleanly the gauge-theoretic character of MBQC. Section~\ref{MBQCgauge}---the most important part of the paper---organizes the material in Sections~\ref{sec:review} and \ref{section:prep} into concepts from gauge theory: a covariant derivative, reference sections, gauge transformations, and fluxes. Section~\ref{equiv} discusses the phenomenological implications of the MBQC gauge \bart{transformations} and shows how the gauge theoretical formulation sheds light on the relation between MBQC and the circuit model.  We close the paper with a discussion of future directions.

\section{Review of measurement-based quantum computation}
\label{sec:review}

Throughout this paper, we write the Pauli operators as $\sigma_x=:X$, $\sigma_y=:Y$, $\sigma_z=:Z$, and use a subscript $j$ to indicate when the Pauli matrix acts on the $j^{\rm th}$ qubit. We let $|\pm\rangle$ be the eigenstates of $X$ with eigenvalues $\pm1$.  We will often refer to these eigenstates using a binary variable $s$ according to $\ket{(-)^s}$, that is $s=0$ stands for $|+\rangle$ and $s=1$ stands for $|-\rangle$. We will also need a notation for eigenstates of $O(\alpha) = (\cos \alpha) X + (\sin \alpha) Y$; they will be denoted with $\ket{\pm_{\alpha}}$. Measurements of $O(\alpha)$ will also be referenced using a binary variable $s$ according to $\ket{\pm_{\alpha}} = \ket{(-)^s_{\phantom{s}\alpha}}$. 

\subsection{The one-dimensional cluster state}
\label{cluster}
The one-dimensional cluster state is a resource state for MBQC, which affords enough computational power to simulate $SU(2)$ gates. For reasons which will become clear in Section~\ref{sec:defflux}, we denote the cluster state as $\cluster$. 

\paragraph{Teleportation and the cluster state } 
The logical processing in MBQC may be understood as a sequence of teleportations. A resource state can be extracted by reordering commuting operations in a sequence of teleportation steps. 
\begin{figure}[t]
\centering
\includegraphics[scale=.23]{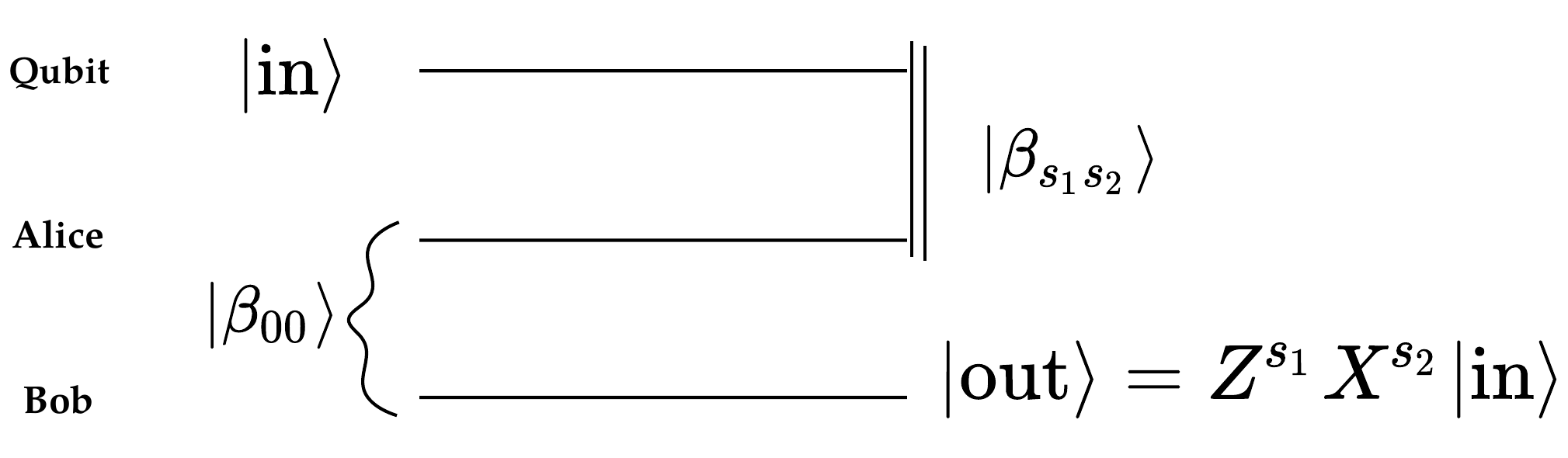}
\quad
\raisebox{.24em}{
\includegraphics[scale=.19]{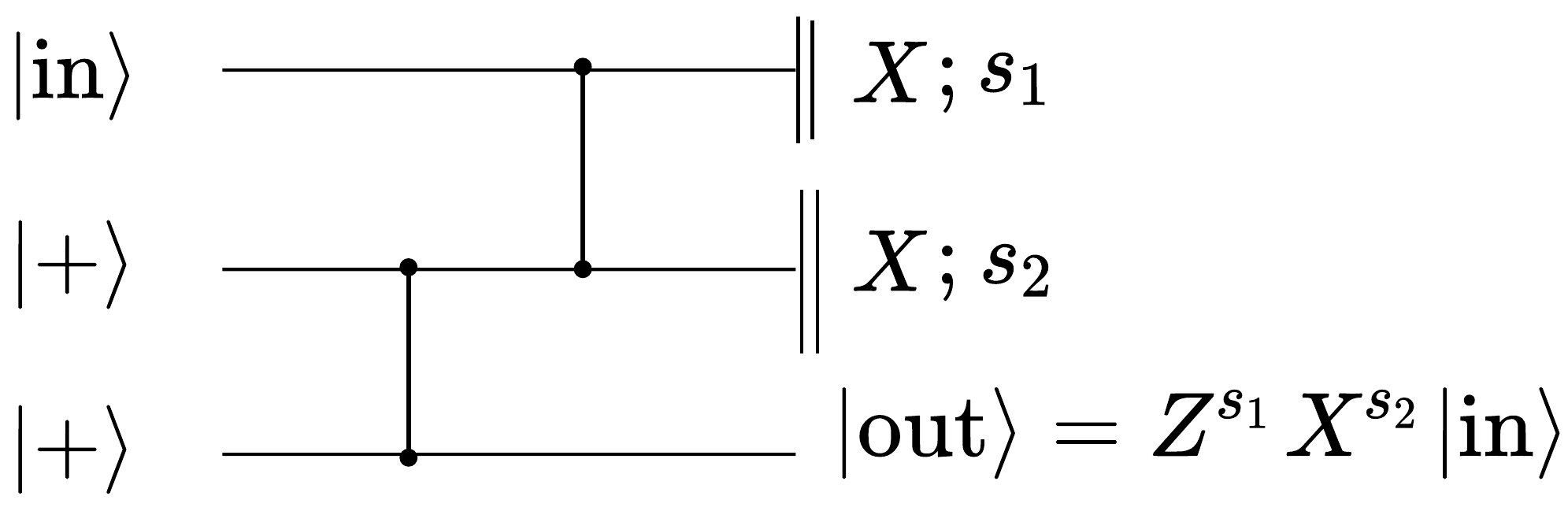}}
\caption{Circuit diagram for quantum teleportation.}
\label{tel} 
\end{figure}
The circuit for quantum teleportation is shown in the left panel of Figure~\ref{tel}. Alice and Bob share a prearranged resource state in the form of a Bell pair $\ket{\beta_{00}} := 2^{-1/2} (\ket{11} +\ket{00})$. To teleport an initial qubit $|{\rm in}\rangle$, Alice makes a measurement in the Bell basis, which comprises states $\ket{\beta_{s_{1}s_{2}}} = X_1^{s_2} Z_1^{s_1} \ket{\beta_{00}}$ with $s_i = 0,1$. (The subscripts in $X_1$ and $Z_1$ refer to the first qubit of the Bell pair.) Because Alice does not control the random outcome of this measurement, the teleported qubit ends up rotated by a known but random byproduct operator 
\begin{align}
V_{s_{2}s_{1}}= X^{s_{2}} Z^{s_{1}} \quad {\rm where}~s_{i} = 0,1.
\label{by}
\end{align}   
In this way, teleportation defines a randomly \textit{fluctuating} circuit. 

This randomness poses a challenge to our ability to perform the most basic computation: to simulate the identity gate on the teleported qubit. Luckily, Bob can undo the effect of the fluctuation because the measurement outcomes $(s_2, s_1)$ are known to Alice, who passes them on to him by classical channels. Once Bob learns the measurement outcome, he can apply the inverse of the byproduct~(\ref{by}) and recover $|{\rm in}\rangle$. What is equally important for the purposes of this paper, the measurement outcomes are related by a symmetry.  The byproduct operators (\ref{by}) form a projective representation of $\mathbb{Z}_2 \times \mathbb{Z}_2$, that is they respect the group operation 
\begin{equation}
(s_2, s_1) + (s_2', s_1') = (s_2 + s_2', s_1 + s_1') \qquad \textrm{(mod~2)}
\label{groupop}
\end{equation}
up to a multiplicative phase. While the group operation~(\ref{groupop}) is not essential in a single instance of quantum teleportation, it makes for a key simplification when multiple teleportations are conducted in sequence. For example, after $N$ sequential teleportations, Bob can apply a single corrective operator to invert
\begin{equation}
B = X^{\left(\sum_{j~{\rm even}}^{2N} s_j\right)} Z^{\left(\sum_{j~{\rm odd}}^{2N-1} s_j\right)}
\end{equation}
and recover the initial state $|{\rm in}\rangle$ up to a phase. We argue that this simplification, which is essential for MBQC, is best understood in the language of gauge theory. The $\mathbb{Z}_2 \times \mathbb{Z}_2$, which permutes the random measurement outcomes in each step, plays the role of a gauge group. 

\begin{figure}[h]
\centering
\includegraphics[scale=.4]{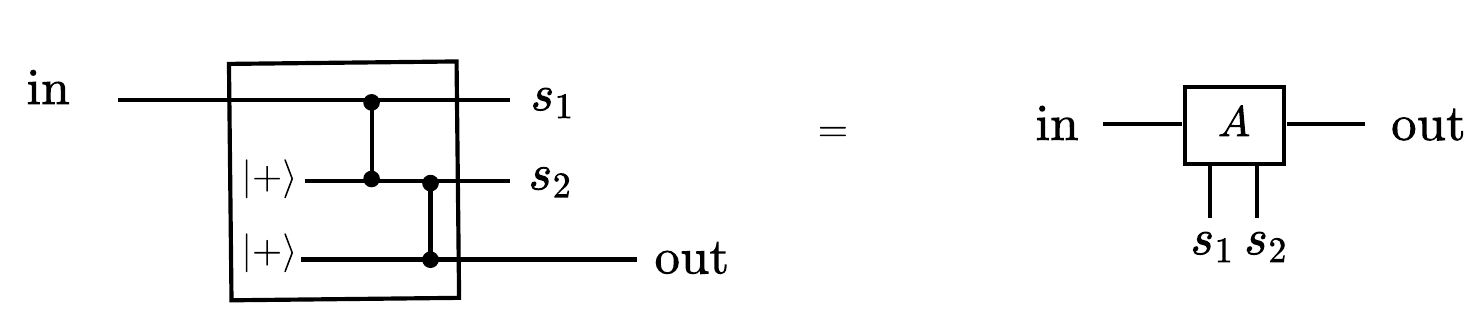}
\caption{MPS tensor from the teleportation circuit. }
\label{MPST}
\end{figure}

A resource state for a single instance of quantum teleportation is the Bell pair $|\beta_{00}\rangle$. A resource state for MBQC must, in a similar fashion, enable multiple teleportations in a sequence. Such a resource is constructed by putting together many copies of the circuit in Figure~\ref{tel} as building blocks. Before the building blocks are assembled, it is useful to redraw them using the cPhase gate
\begin{figure} [h]
\centering
\includegraphics[scale=.4]{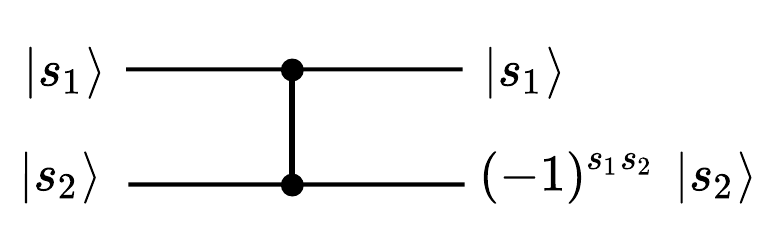}
\label{cphase} 
\end{figure}
\begin{equation}
{\rm cPhase}(i,j) := \mathbb{I}_{i,j} - \frac{(\mathbb{I}_i-Z_i)(\mathbb{I}_j-Z_j)}{2} 
\label{cp}
\end{equation} 
as is done in the right panel of Figure~\ref{tel}. In equation~(\ref{cp}), $i$ and $j$ label the two qubits on which cPhase acts. The redrawing reveals that Alice's measurement in the $|\beta_{s_1 s_2}\rangle$ basis is equivalent to measurements in the $X$-basis on the state shown in Figure~\ref{MPST}. 

\begin{figure}
        \centering
        \includegraphics[width=0.420\textwidth]{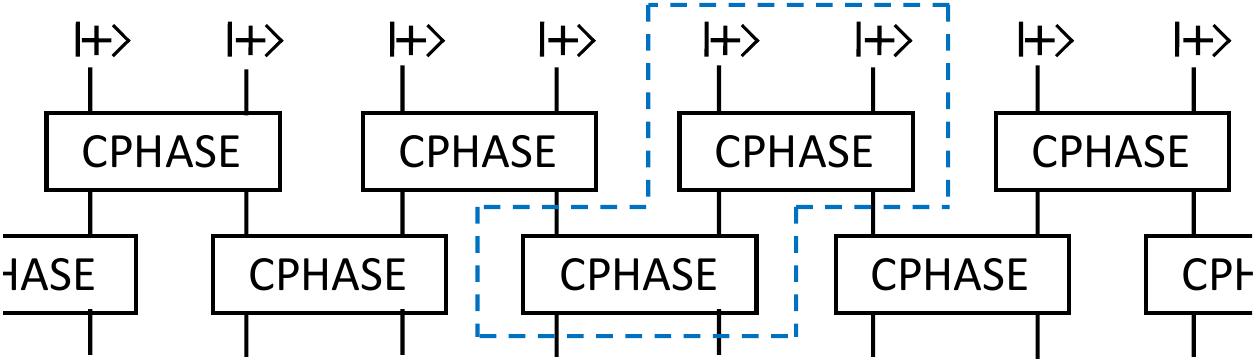}
                \hfill
       \raisebox{0.04\textwidth}{\includegraphics[width=0.30\textwidth]{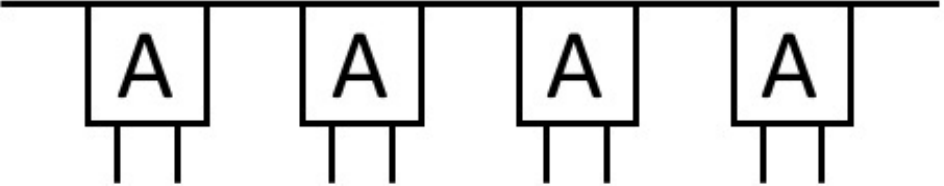}}      
               \hfill
        \raisebox{0.028\textwidth}{\includegraphics[width=0.166\textwidth]{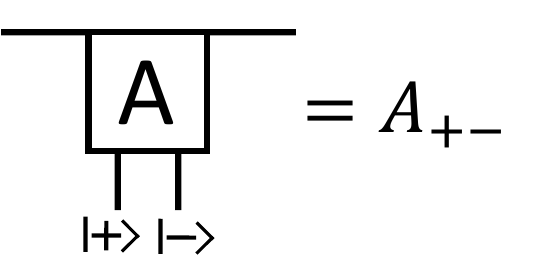}}
        \caption{The cluster state, as given by equation~(\ref{clusterdef}) (left) and in its MPS form (middle). The highlighted part of the left panel is one matrix of the MPS description (right); it covers one `block.'}
        \label{fig:clusterbasic}
\end{figure}

To enable multiple teleportations to proceed in a sequence, we concatenate copies of Figure~\ref{MPST}, identifying the $\langle {\rm out}|$ leg of one copy with the $| {\rm in}\rangle$ leg of the next copy. The state obtained this way is a matrix product state (MPS) and Figure~\ref{MPST} represents its MPS tensor. This MPS state---the one-dimensional cluster state $\cluster$---is shown in Figure~\ref{fig:clusterbasic}. 

\paragraph{The cluster state as a matrix product state} 

It is important to distinguish two Hilbert spaces involved in the MPS tensor. Alice's measurement of the indices marked $s_2 s_1$ in Figure~\ref{MPST} effects quantum teleportation---a linear map from $|{\rm in}\rangle$ to $|{\rm out}\rangle$. As such, the MPS tensor is an $(s_2, s_1)$-dependent linear map $A_{s_2 s_1}$ with matrix elements $\langle {\rm out}| A_{s_2 s_1} | {\rm in} \rangle$. In discussions of MPS states, it is standard terminology to call the Hilbert space spanned by $|{\rm in}\rangle$ and $|{\rm out}\rangle$ the virtual space or correlation space, while the Hilbert space denoted by the subscript $\cdot_{s_2 s_1}$ is referred to as the physical space. Virtual space is called virtual because, when the $\langle {\rm out}|$ leg of one MPS tensor is joined with the $| {\rm in}\rangle$ leg of the next (when two teleportations are conducted in a sequence), we lose direct physical access to that degree of freedom. The physical space is called physical because it is always physically accessible; it is the space where Alice makes her measurements. 

Our discussion should make clear that the MPS tensor $A_{s_2 s_1}$ in the $X$-basis equals the byproduct $V_{s_2 s_1}$ from equation~(\ref{by}). 
The cluster state on a ring of $2N$ qubits is obtained by contracting $N$ copies of this MPS tensor:
\begin{equation}
\!\!\!\!\!\!\cluster = \mathcal{N}
\sum_{(s_{2j},s_{2j-1}) \in \mathbb{Z}_2 \times \mathbb{Z}_2 } 
{\rm Tr} \big(A_{s_{2N},s_{2N-1}} \ldots  A_{s_4,s_3}A_{s_2,s_1}\big) 
|s_{2N},s_{2N-1}\rangle \ldots |s_4,s_3\rangle |s_2,s_1\rangle 
\label{mpsring}
\end{equation}
Each $|s_{2j},s_{2j-1}\rangle$ is a two-qubit state taken from $\{|\!++\rangle, |\!+-\rangle, |\!-+\rangle, |\!--\rangle\}$, labeled with an element $(s_{2j},s_{sj-1}) \in \mathbb{Z}_2 \times \mathbb{Z}_2$.   We use the generic notation $A_{(s_{2j},s_{2j-1})}$ for the MPS tensor rather than $V_{(s_{2j},s_{2j-1})}$ since we will later be interested in evaluating the same tensor in a different basis. Here and in similar formulas below the factor $\mathcal{N}$ ensures a proper normalization of the state.

Throughout this paper, the notation `$|1\rangle$' shall stand for the 1D cluster state, not for a computational basis state. The reason for this notation will become clear later. In short, the cluster state is identified as an eigenstate of a flux observable with eigenvalue 1.

For a chain-like cluster state, the traces in the amplitudes are replaced by analogous expressions, which depend on a choice of boundary conditions at endpoints:
\begin{equation}
{\rm Tr} \big(A_{s_{2N},s_{2N-1}} \ldots  A_{s_4,s_3}A_{s_2,s_1}\big) 
\,\,\longrightarrow\,\,
\langle {\rm out} |A_{s_{2N},s_{2N-1}} \ldots  A_{s_4,s_3}A_{s_2,s_1} | {\rm in} \rangle
\label{chaina}
\end{equation}

\paragraph{A glimpse of a global structure.}
Equation~(\ref{mpsring}) offers a first sighting of the kind of global structure, which we seek to identify. This comes in the form of a global selection rule, which determines which combinations of physical spins do / do not overlap with the ring-like cluster state. 

The amplitudes in (\ref{mpsring}) involve products of projective representation matrices of $\mathbb{Z}_2 \times \mathbb{Z}_2$. We can use the group multiplication $A_{(s_{2j+2},s_{2j+1})} A_{(s_{2j},s_{2j-1})} \propto A_{(s_{2j+2},s_{2j+1})+(s_{2j},s_{2j-1})}$ to simplify them:
\begin{equation}
\!\cluster=
\mathcal{N}\sum_{(s_{2j},s_{2j-1}) \in \mathbb{Z}_2 \times \mathbb{Z}_2} 
\textrm{(phase)} \times ({\rm Tr} A_{(s_{2N},s_{2N-1}) +  \ldots (s_4,s_3)+ (s_2,s_1)}\,
|s_{2N},s_{2N-1}\rangle \ldots |s_4,s_3\rangle |s_2,s_1\rangle
\label{mpsring2}
\end{equation}
Since every $A_{(\cdot)}$ except $A_{++}$ is traceless, the wavefunction vanishes unless 
\begin{equation}
(s_{2N},s_{2N-1}) \ldots (s_4,s_3)+ (s_2,s_1) = (0,0) \in \mathbb{Z}_2 \times \mathbb{Z}_2
\label{clusterholonomy}
\end{equation}
Component-wise, we thus have
\begin{equation}
a\equiv  s_1 + s_3 + \ldots + s_{2N-1} = 0 \qquad {\rm and} \qquad b\equiv s_2 + s_4 + \ldots + s_{2N} = 0 \quad \textrm{(mod 2)}
\label{clusterselect}
\end{equation}

We will recast the selection rule (\ref{clusterholonomy}-\ref{clusterselect}) as a holonomy $(a,b)$ of a $\mathbb{Z}_2 \times \mathbb{Z}_2$ gauge field. In this view, the cluster state on a ring represents a gauge theory state with trivial $\mathbb{Z}_2 \times \mathbb{Z}_2$ flux. Equivalently, the Wilson line around the cluster evaluates to $++$, which is the identity element of the group. This is a first reason why we denote the cluster state as $\cluster$, though a more complete explanation will be given in Section~\ref{sec:defflux}. 

Readers familiar with MBQC will recognize the meaning of this discussion. When we measure $\cluster$ in the $|\pm\rangle$ basis, we effectively simulate the identity gate. This simulation happens to have a fully deterministic outcome---reflecting the fact that $\cluster$ is a state of definite $++$ (identity) flux.

\paragraph{The cluster state is short ranged entangled}

Figure~\ref{MPST} shows that the MPS tensor can be constructed by acting with two-qubit entangling gates~(\ref{cp}) on a product state. Naturally, the same can be said about the full MPS state. The left panel of Figure~\ref{fig:clusterbasic} shows (a segment of) the cluster state, constructed by repeated applications of the cPhase gate on a product state. As an equation, we have:
\begin{equation}
\cluster := \prod_{i=1}^{2N-1~{\rm or}~2N} \!\!\!\!\!\!\!\text{cPhase}(i,i+1) 
\left(\bigotimes_{j=1}^{2N} |+\rangle_j\right)
\label{clusterdef}
\end{equation}
The $(2N)^{\rm th}$ gate $\text{cPhase}(2N,2N+1)\equiv \text{cPhase}(2N,1)$ is only applied on a ring, where we define $2N+1 \equiv 1$. Note that all cPhases commute because they are simultaneously diagonal in the $Z$-eigenbasis. 

As seen in Figure~\ref{fig:clusterbasic}, the preparation of the cluster from the product state $\otimes_j |+\rangle_j$ can be achieved in two time steps: the application of the cPhases$(i,i+1)$ with $i$ odd, and with $i$ even. Because the cluster state can be prepared by a circuit of finite depth (two), it is short range-entangled. This demonstrates that using the cluster state for quantum computation does not introduce hidden costs associated with the preparation of the initial state. 

\paragraph{The cluster state as a stabilizer state.}
Equation~(\ref{clusterdef}) reveals that $\cluster$ is a stabilizer state \cite{GomaSM,Goma2SM}. Because every $X_j$ stabilizes $\otimes_j |+\rangle_j$, conjugating $X_j$ by the cPhases in (\ref{clusterdef}) gives a stabilizing operator for $\cluster$. Using the identity ${\rm cPhase}(i,j) X_i\, {\rm cPhase}(i,j) = X_i Z_j$, we find the following stabilizers:
\begin{equation}
K_j \cluster = \cluster \qquad {\rm with}~~K_j := Z_{j+1} X_j Z_{j-1} 
\label{defstab}
\end{equation}
On a ring, these equations completely determine $\cluster$. On a chain indexed by $1 \leq j \leq 2N$, equation~(\ref{defstab}) makes no sense for $j = 1$ or $2N$. In that context, in addition to (\ref{defstab}), the chain-like cluster state is also stabilized by two special operators $Z_2 X_1$ and $X_{2N} Z_{2N-1}$. The $K_j$ with $2 \leq j \leq 2N-1$, along with the two special stabilizers at endpoints, completely determine the cluster state on a chain. For future use, we remind the reader that each $K_j$ squares to the identity. 

The stabilizer conditions~(\ref{defstab}) translate the entanglement structure of the cluster state into a set of algebraic constraints, which can be solved explicitly by summing over orbits of the stabilizers:
\begin{align}
\label{Kcon}
\cluster\propto \prod_{i}  \left(\frac{1+K_{i}}{2}\right)\left(\bigotimes_{j=1}^{2N} |+\rangle_j\right)
\end{align} 
The structure of these constraints as well as their solution mirrors the structure of gauge constraints, which is why we will be able to embed the cluster state in a gauge theory.\footnote{In high energy theory, one usually defines the gauge theory Hilbert space as a quotient $\mathcal{H} = \tilde{\mathcal{H}} / G$, where $G$ \bart{is the gauge group.} If the gauge theory is fundamental, the pre-quotient Hilbert space $\tilde{\mathcal{H}}$ is an auxiliary construct and $G$ {\it physically} acts as the identity. The situation is \bart{more intricate} when the gauge theory is not fundamental but a truncation of another, more fundamental theory whose Hilbert space is $\tilde{\mathcal{H}}$. This situation, more common in condensed matter theory, arises for example when an energy gap separates $G$-invariant states from other states in $\tilde{\mathcal{H}}$. Our discussion assumes the latter setup. Specifically, the individual qubits that make up the cluster state live in $\tilde{\mathcal{H}}$ and generators of gauge \bart{transformations} $K_i$ act on them as physical operators. We will refer to the Hilbert space of qubits as `physical' and the quotiented space as `flux space'.}   
The stabilizer conditions play an important role in mitigating the randomness of measurement outcomes in MBQC. Much of our paper is devoted to understanding this feature of MBQC in terms of \bart{gauge transformations}. For a preview of this interpretation, observe that for any operator $O$, conjugation by a stabilizer does not affect expectation values taken in $\cluster$:
\begin{equation}\label{KOK}
\sterclu K_j O K_j \cluster = \sterclu O \cluster
\end{equation}
This means that, so long as we work with the cluster state $\cluster$, it is redundant and unnecessary to distinguish $O$ from $K_j O K_j$. For a technician operating on $\cluster$, the difference between $O$ and $K_j O K_j$ is a redundancy of description\bart{, which can be conceptualized as a change of gauge.}

Because the stabilizers $K_j$ commute with one another, the full stabilizer group of the cluster state is isomorphic to $\mathbb{Z}_2^{2N}$.   In fixing a `stabilizer gauge,' one independent binary choice---to apply or not to apply $K_j$ (as in \eqref{Kcon})---is available at every qubit $j$.   The independence of these local choices\bart{---one $\mathbb{Z}_2$ per qubit---agrees with the conventional understanding of the origin of gauge as a local description of a global concept.} In fact, the $K_j$s which act on even and odd sites can be meaningfully distinguished from one another.  We will see that this implies the cluster state \bart{is preserved by a local $\mathbb{Z}_2 \times \mathbb{Z}_2$, and that the gauge character of this group is an essential component of MBQC.}

\subsection{MBQC}
Here we review MBQC on a chain, following the treatment of \cite{Eis2} in the MPS formalism. In Section~\ref{BA} we introduce the idea of a logical quantum register propagating in the MPS virtual (or correlation) space, first for the special instance where all measurements are performed in the local $X$-basis. In Section~\ref{MBTO} we introduce more general local measurement bases, leading to more MBQC-simulated quantum gates. In Section~\ref{sec:output} we describe how to extract the computational output from the measurement record. 

\subsubsection{Basic idea}
\label{BA}
To begin, we write down the cluster state on a chain as defined by equations~(\ref{mpsring}) and (\ref{chaina}):
$$
\cluster = \mathcal{N}
\sum_{s_j \in \mathbb{Z}_2 } 
\Big( \langle {\rm out} | A_{s_{2N},s_{2N-1}} \ldots A_{s_4,s_3} A_{s_2,s_1} | {\rm in} \rangle \Big)
|s_{2N},s_{2N-1}\rangle \ldots |s_4,s_3\rangle |s_2,s_1\rangle.
$$
At present, we consider the above expansion in the local $X$-basis only. This restriction will subsequently be lifted. One way to interpret this formula is as a sum over trajectories, which an input state $|{\rm in} \rangle$ follows under successive applications of operators $A_{s_{2j},s_{2j-1}}$. In a laboratory, an experimenter can (partly) control these trajectories by projecting the cluster state onto chosen states of successive qubits.  To wit, measuring the first two qubits in the $\langle s_2,s_1|$ basis applies the gate $A_{s_2,s_1} = V_{s_2,s_1}$:
\begin{equation}
\langle s_2,s_1\! \cluster =
\mathcal{N}'
\sum_{s_j \in \mathbb{Z}_2 } 
\Big( \langle {\rm out} | A_{s_{2N},s_{2N-1}} \ldots A_{s_4,s_3} V_{s_2,s_1} | {\rm in} \rangle \Big)
|s_{2N},s_{2N-1}\rangle \ldots |s_4,s_3\rangle
\end{equation}
Note that the matrix $V_{s_2,s_1}$, which is left inside the amplitude after the $\langle s_2,s_1|$ projection, is not an MPS tensor anymore because its physical indices are no longer summed over. This is why we denote it as $V_{s_2,s_1}$ and not $A_{s_2,s_1}$; see the comment above equation~(\ref{mpsring}).
A sequence of projections determines a longer trajectory:
\begin{equation}
\ldots
\,\longleftarrow\, V_{s_6,s_5} V_{s_4,s_3} V_{s_2,s_1} |{\rm in}\rangle
\,\longleftarrow\, V_{s_4,s_3}V_{s_2,s_1} |{\rm in}\rangle
\,\longleftarrow\, V_{s_2,s_1} |{\rm in}\rangle 
\,\longleftarrow\, |{\rm in}\rangle
\label{trajectory}
\end{equation}
MBQC uses such trajectories to realize quantum computation. Indeed, (\ref{trajectory}) can be viewed as a simulation of a computation in the circuit model, in which the applied gates are of the form~(\ref{by}). The lab technician, who projects $\cluster$ onto chosen single qubit states, is a quantum computer programmer. In the final step of this quantum computation, the last state on the trajectory
\begin{equation}
V_{s_{2N},s_{2N-1}} \ldots V_{s_4,s_3} V_{s_2,s_1} | {\rm in} \rangle = 
{\rm (phase)} \times V_{(s_{2N},s_{2N-1}) + \ldots + (s_2,s_1)} | {\rm in} \rangle
\end{equation}
is measured in some basis $\langle {\rm out}|$ to produce one classical bit of computational output. 

The above description leaves out two important aspects of MBQC. The first one concerns the set of quantum gates, which MBQC can simulate. Thus far, we have only drawn gates from~(\ref{by}), but for realizing general $SU(2)$ transformations a larger gate set is necessary. This will require a generalization of the quantum teleportation protocol, in which the measurement basis is modified.   The second wanting point, which is conceptually more important, is about control over measurement outcomes. The preceding discussion pretended that the technician could project $\cluster$ onto chosen qubit states at will, but of course every quantum measurement is probabilistic. When the technician does not obtain her planned outcome, the virtual qubit in (\ref{trajectory}) takes an unplanned turn. For a complete definition of MBQC, we must give a protocol to address both of those outstanding points. We do so presently.

\subsubsection{Measurement basis and temporal order}
\label{MBTO}
In order to simulate an arbitrary $SU(2)$ gate, MBQC generalizes the teleportation protocol by measuring the cluster state in a rotated basis on the $XY$ plane.
To prevent the randomness of measurement outcomes from contaminating the quantum computation, future measurement angles must be adapted based on past  outcomes. These two features can be achieved by measurements in a temporally ordered basis, given by
\begin{equation}\label{orot}
O\big((-)^{q_{i}}\alpha_{i}\big) := 
\cos (\alpha_{i})X + \sin\big((-)^{q_{i}}\alpha_{i}\big) Y ,
\end{equation}
where 
\begin{equation}
\label{CPR1}
q_i =  \, s_{i-1} + s_{i-3} + \ldots + \big( s_2~{\rm or}~s_1 \big). 
\end{equation}
Here $s_{i}$ denotes the measurement outcome for measuring along $(-)^{q_i} \alpha_{i}$. 
The parameters $q_{i}$ tell the experimenter measuring the $i^{\rm th}$ qubit to adapt the sign of $\alpha_{i}$ based on results of previous measurements. Eq.~(\ref{CPR1}) is one of the two \textit{classical side processing relations} of MBQC. It governs the adaptation of measurement bases to account for the randomness of earlier measurement outcomes.

To understand the adapted basis \eqref{CPR1} in the simplest setting, consider measurements on a single {\it block} of two sites. (We will refer to consecutive pairs of spins as `blocks' throughout the paper; cf. Figure~\ref{fig:clusterbasic}.) Let us denote the projection of the MPS tensor onto $O(\alpha_{i})$-eigenstates with $A_{s_{2}s_{1}}\!(\alpha_{2},\alpha_{1})$.   Equation \eqref{CPR1} instructs us to flip the sign of $\alpha_{2}$ based on the result of measuring the first spin along $\alpha_{1}$ because $q_2 = s_1$.
Doing so results in the following operators acting on the virtual space:
\begin{equation}\label{arot}
\begin{array}{rcl}
A_{++}(\alpha_2, \alpha_1) &=& V_{++}\, U\\
A_{-+}(\alpha_2, \alpha_1) & = & V_{-+} \, U\\
A_{+-}(-\alpha_2, \alpha_1) & = & V_{+-} \, U\\
A_{--}(-\alpha_2, \alpha_1) & = & V_{--}\, U
\end{array}
\end{equation}
where
\begin{equation}
U = \exp\left(-i\frac{\alpha_2}{2}X\right) \exp\left(-i\frac{\alpha_1}{2}Z\right)
=A_{++}(\alpha_2, \alpha_1) 
\label{defu}
\end{equation}
and $V_{s_{2}s_{1}}$ was defined in \eqref{by}. These equations generalize  teleportation, in the sense that the `teleported' input state arrives rotated by a unitary transformation $U$, up to a random byproduct operator $V_{s_{2}s_{1}}$.  Ordinary teleportation corresponds to $U=1$ and is recovered when $\alpha_{i}=0$.

Now consider a cluster chain with $N$ blocks, that is $2N$ spins. We find in eq.~(\ref{arot}) that for all four combinations of measurement outcomes, the unitary $U$ is the same, and the evolution of the virtual system differs only by the random byproduct operator. The byproduct operators $V$ are all of Pauli type.

These random byproduct operators, one produced by each measurement, need to be removed from the computation. This is achieved by propagating them forward in time, through the not yet implemented part of the computation, and past the logical readout measurements. What happens after logical readout doesn't affect the computational output.

Inspecting equation~(\ref{defu}), we find:
\begin{equation}
A_{++}(\alpha_2,\alpha_1)X = XA_{++}(\alpha_2,-\alpha_1)
\qquad {\rm and} \qquad
A_{++}(\alpha_2,\alpha_1)Z = ZA_{++}(-\alpha_2,\alpha_1).
\end{equation}
The forward-propagation of the byproduct operators flips angles of rotations on virtual space, and those flips have to be compensated by flips of the measurement angle, cf. Eq.~(\ref{orot}). The binary observables $q_i$ account for this. With eqs.~(\ref{by}) and (\ref{defu}) we verify that the sign flip of the measurement angle on an even site depends on the parity of all measurement outcomes on earlier odd sites, and the sign flip of the measurement angle on an odd site on the parity of the measurement outcomes of the earlier even sites. This is the content of the classical processing relation~(\ref{CPR1}).

Denoting the MPS tensor in the adapted basis by
\begin{equation}
 A_{s_{i+1},s_{i};q_{i+1},q_{i}}:=A_{s_{i+1},s_{1}}((-1)^{q_{i+1}} \alpha_{2N},(-1)^{q_{i}}\alpha_{2N-1}) ,
\end{equation}
the result of the byproduct propagation can be summarized by
\begin{equation}
 A_{s_{2N},s_{2N-1};q_{2N},q_{2N-1}} \ldots  A_{s_2,s_1;q_2,q_1} = 
 B\, U_{\rm total},
\label{burelation}
\end{equation}
where
\begin{align}
B & = V_{s_{2N},V_{2N-1}} \ldots V_{s_4,s_3} V_{s_2,s_1} = {\rm (phase)} \times V_{(s_{2N},s_{2N-1}) + \ldots + (s_2,s_1)}
\label{defb},\\
U_{\rm total} & = A_{++} (\alpha_{2N}, \alpha_{2N-1}) \ldots A_{++}(\alpha_4, \alpha_3) A_{++} (\alpha_2, \alpha_1).
\label{defutotal}
\end{align}
We now recognize that equation~(\ref{orot}) is designed to render (\ref{defutotal})---so that the adapted measurements always simulate the same unitary transformation $U_{\rm total}$. 

Equation \eqref{burelation} is the most important result of the MBQC protocol.   It shows that when the cluster state is measured in the adapted basis, the `teleported' qubit arrives rotated by $U_{\rm total}$, up to a known byproduct operator $B$ for which the recipient must correct.  In Section~\ref{MathPin} we explain that $B$ functions as a $\mathbb{Z}_2 \times \mathbb{Z}_2$ Wilson line.

\paragraph{Simulating a general unitary}
Finally, observe that a general $SU(2)$ gate can be generated by measuring four qubits:
\begin{equation}
A_{++}(0, \alpha_3) A_{++} (\alpha_2, \alpha_1) = 
\exp\left(-i\frac{\alpha_3}{2}Z\right)
\exp\left(-i\frac{\alpha_2}{2}X\right) \exp\left(-i\frac{\alpha_1}{2}Z\right)
\label{mbqcminimum}
\end{equation}
This expression parametrizes the $SU(2)$ manifold with Euler angles $\alpha_{1,2,3}$. We conclude that measurements of $\cluster$ in the $O(\alpha)$ eigenbasis can simulate every $SU(2)$ gate on virtual space so long as the effect of byproduct operators---an imprint of the randomness of measurement outcomes---can be properly remedied. 

\bartek{The four qubits in equation~(\ref{mbqcminimum}) represent the minimal number to effect an $SU(2)$ computation. But the same computation can be realized in MBQC in multiple ways, both in terms of the length of the chain and the specific set of measurement outcomes.  All MBQC realizations of the same computation are described by equation~(\ref{burelation}), with $B$ interpreted as a $\mathbb{Z}_2 \times \mathbb{Z}_2$ Wilson line. Our treatment of MBQC as a gauge theory unifies all MBQC realizations of the same computation.}

\subsubsection{Computational output for MBQC on a chain}
\label{sec:output}
After qubits 1 through $2N$ of a chain-like cluster state have been measured in the adapted basis, the experimenter gets a probability distribution, which depends on the boundary condition $\langle {\rm out}|$ at the other end. Instead of a preordained boundary condition, we will follow the logic of equation~(\ref{Kig}) and treat the choice of $\langle {\rm out}|$ as one final projective measurement, applied to the $(2N+1)^{\rm st}$ qubit in the state:
\begin{equation}
 {\rm (phase) } \times B \, U |{\rm in}\rangle
\label{afterqubits}
\end{equation}
The $B$ and $U$ are given in~(\ref{defb}-\ref{defutotal}) and we drop the subscript `total' from $U$.

In the circuit model of quantum computing, the application of a gate $U$ to an initial state $|{\rm in}\rangle$ is followed by a measurement of $U |{\rm in}\rangle$ in some orthogonal basis, for example $\langle \pm|$. The computational output is sampled from the resulting probability distribution $p_\pm = | \langle \pm | U | {\rm in} \rangle |^2$.  

 In MBQC, we mirror that last step by measuring state~(\ref{afterqubits}) in an orthogonal basis $\langle {\rm out}_\pm|$, with $\langle \pm|$ being a canonical choice. The hitherto unfixed choice of boundary condition $\langle {\rm out}|$ at the end of the chain represents the choice of basis for the final measurement, whose outcome serves as the computational output.

In equation~(\ref{afterqubits}), the randomness of prior measurement outcomes remains visible in the byproduct operator $B$. For a computational output that is free of randomness, we again adjust the measurement basis and project state~(\ref{afterqubits}) onto $\langle {\rm out}_\pm | B$ instead of $\langle {\rm out}_\pm |$. This works because $B^2 = \pm 1$ implies:
\begin{equation}
\langle {\rm out}_\pm | B \, {\rm (phase)} \times B \, U |{\rm in}\rangle  
= \textrm{(phase)} \times \langle {\rm out}_\pm | U | {\rm in} \rangle
\label{whattomeasure}
\end{equation}

When the final measurement is done in the $X$-eigenbasis $\langle {\rm out}_\pm| = \langle \pm|$, the adjustment $\langle {\rm out}_\pm| \to \langle {\rm out}_\pm| B$ is particularly simple. Let us write the byproduct $B$ in terms of the `holonomy' variables $(a,b)$ from \eqref{clusterselect}:
\begin{equation}
\label{BZX}
B = X^{b} Z^{\tilde{a}} \qquad {\rm with}~~
\tilde{a}=\sum^{2N-1}_{j~{\rm odd}} s_j \quad {\rm and} \quad b=\sum^{2N}_{j~{\rm even}} s_j
\end{equation}
We ignore the possible phase because it will drop out of the reported probability distribution $p_\pm = | \langle {\rm out}_\pm | U | {\rm in} \rangle |^2$. Letting $\langle {\rm out}_\pm| :=\bra{s_{2N+1}}$, we see that the byproduct transforms the final projection basis as:
\begin{align}
\bra{s_{2N+1} } B = \bra{s_{2N+1} + \tilde{a} } := \bra{a}
\end{align}
Accordingly, we define the computational output as:
\begin{equation}
a := s_{2N+1} + \tilde{a} = s_1 + s_3 + \ldots + s_{2N-1} + s_{2N+1}
\label{chainoutput}
\end{equation}
Eq.~(\ref{chainoutput}) is the second classical processing relation of MBQC. Together with eq.~(\ref{CPR1}), which governs the adaptation of measurement bases, it is responsible for compensating the randomness inherent in the quantum measurements driving MBQC.

The probability $p(a)$ of observing the output $a = 0,1$ is exactly equal to the probability $p(a) = |\braket{a|U|\rm{in}}|^2$ for the corresponding quantum circuit realizing the unitary $U$ and finding $a$ upon readout.

Equation~(\ref{chainoutput}) puts all $2N+1$ measurements on equal footing. Here, again, a global structure that underlies MBQC becomes evident. We have advertised in the Introduction, and we argue in Section~\ref{MathPin}, that the MBQC computational output is (one component of) a Wilson line of a $\mathbb{Z}_2 \times \mathbb{Z}_2$ gauge field. 

\section{Preparations for the MBQC gauge theory}
\label{section:prep}
To set the stage for the presentation of the MBQC gauge theory in Section~\ref{MBQCgauge}, we introduce two additional ingredients in this section: a group-valued label for the measurement outcomes with respect to the adapted basis \eqref{orot}, and an algorithm for performing MBQC on a ring.  

\subsection{$\mathbb{Z}_2 \times \mathbb{Z}_2 $ action and MPS tensor symmetries} 
We begin by reviewing the relevant representations of $\mathbb{Z}_2 \times \mathbb{Z}_2 $ on the physical Hilbert space and the transformation laws of the MPS tensor under these symmetries \cite{SPT2,Schuch}.  There are two projective representations of  $\mathbb{Z}_2 \times \mathbb{Z}_2 $, which we refer to as $l(g)$ and $r(g)$.  Using the multiplicative mod 2 notation $g=(\pm,\pm) \in \mathbb{Z}_2 \times \mathbb{Z}_2 $, their action on each block of physical qubits are:
\begin{align}
l(++)&= 1\otimes 1,\quad l(+-) = X_2 \otimes Z_1, 
\quad l(-+) = Z_2 \otimes 1 ,\quad l(--)= Z_2 X_2 \otimes Z_1
\label{lgprojective}\\
r(++)&=1 \otimes 1, \quad r(+-) = 1\otimes Z_1,\quad r(-+)=  Z_2 \otimes X_1, \quad r(--) = Z_2 \otimes X_1 Z_1
\label{rgprojective}
\end{align} 
We refer to these as the left and right action of $\mathbb{Z}_2 \times \mathbb{Z}_2 $.\footnote{\bart{In the context of SPT order, the physical meaning of (\ref{lgprojective}) and (\ref{rgprojective}) is that, in the presence of boundaries, they generate {\it edge modes} or {\it boundary modes}, i.e. excitations localized near the boundaries, which respect the SPT symmetry. For a detailed discussion, see e.g.~\cite{Yoshida}.}}
The product 
\begin{equation}\label{LinRep}
u(g):= l(g)r(g)
\end{equation}
generates a linear representation  $\mathbb{Z}_2 \times \mathbb{Z}_2 $ which acts by $X$ on even and odd sites:
\begin{equation}
u(++)=1 \otimes 1, \quad u(+-) = X_2 \otimes 1,\quad u(-+)=  1 \otimes X_1, \quad u(--) = X_2 \otimes X_1
\label{uglinear}
\end{equation} 

\begin{figure}[t]
\centering 
\includegraphics[scale=.4]{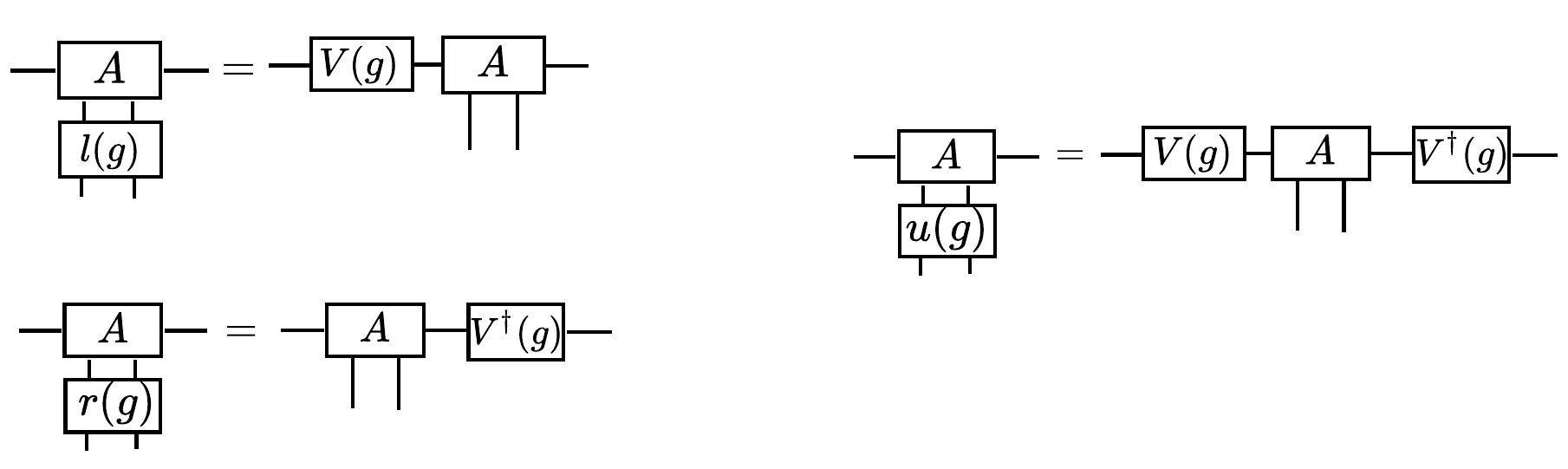} 
\caption{ The left panels show the equivariant transformation law of the cluster MPS tensor  under the left $l(g)$ and right $r(g)$ action of  $\mathbb{Z}_2 \times \mathbb{Z}_2$. Composing these gives the transformation under $u(g)=l(g)r(g)$. More generally, analogous group actions and transformation laws apply to the fixed point MPS of any SPT phase protected by an onsite $G$ symmetry \cite{Pgarcia, Schuch}.} 
\label{equivar}
\end{figure}

The MPS tensor satisfies a set of equivariant transformation laws under (\ref{lgprojective}-\ref{uglinear}).  These are illustrated graphically in Figure~\ref{equivar}.  They relate the action of $\mathbb{Z}_2 \times \mathbb{Z}_2$ on the physical and virtual degrees of freedom, giving the experimentalist indirect access to correlation space via manipulations on physical qubits.  

\subsection{Group labels for measurement outcomes }
We can use the left action $l(\mathbb{Z}_2 \times \mathbb{Z}_2)$ to  assign a group label to the measurement outcomes in each block.  For simplicity let us begin with a chain containing only a single block.  $l(g)$ acts on the $O(\alpha_{i})$ eigenstates according to the commutation relations:
\begin{align}
\label{XOX}
X_{i}\, O(\alpha_{i}) &= O(-\alpha_{i})\, X_{i}\nn
Z_{i}\, O(\alpha_{i}) &= -O(\alpha_{i})\, Z_{i}
\end{align}  
In effect, $Z_{i}$ flips the measurement outcomes $s_{i}$. $X_{i}$, on the other hand, flips the sign of $\alpha_{i}$ but leaves $s_{i}$ invariant. Because equation~(\ref{lgprojective}) features $X_2$ only in the combination $X_2 \otimes Z_1$, we see that $\alpha_2$ gets flipped only when $s_1 = 1$. Equation~(\ref{orot}) encodes this fact in the classical processing relation $q_2 = s_1$.  

Under the action of this symmetry, the locally adapted basis
\begin{equation}
|+_{\alpha_2}\rangle \otimes |+_{\alpha_1}\rangle
\quad {\rm and} \quad
|+_{(-\alpha_2)}\rangle \otimes |-_{\alpha_1}\rangle
\quad {\rm and} \quad
|-_{\alpha_2}\rangle \otimes |+_{\alpha_1}\rangle
\quad {\rm and} \quad
|-_{(-\alpha_2)}\rangle \otimes |-_{\alpha_1}\rangle
\label{localmbqcbasis}
\end{equation}  
forms an orbit under $\mathbb{Z}_2 \times \mathbb{Z}_2$. Therefore, we can label the basis vectors by group elements using $\ket{+_{\alpha_{2}}}\otimes \ket{+_{\alpha_{1}}} $ as a reference:
\begin{align}
\label{ga} 
\ket{g}= l(g)\ket{+_{\alpha_{2}}}\otimes \ket{+_{\alpha_{1}}}, \quad {\rm where}~~g \in  \mathbb{Z}_2 \times \mathbb{Z}_2
\end{align} 
Thus, the measurement outcomes can be identified with elements of $\mathbb{Z}_2 \times \mathbb{Z}_2$.   We use the same group label to denote the MPS tensor components $\eqref{defu}$ with respect to the adapted basis \eqref{ga}:
\begin{align}\label{Ag2}
    A_{g}:= A_{g} ((-)^{s_{1}} \alpha_{2},\alpha_{1}),
\end{align}
where we will occasionally drop the argument of $A_{g}$ when there is no risk of confusion.

\begin{figure}[h]
\centering 
\includegraphics[scale=.6]{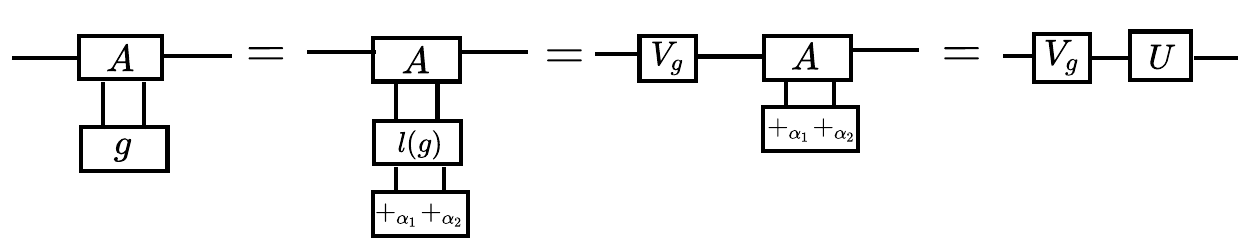}
\caption{The MPS symmetry transformation under $l(g)$ can be applied to decompose each measured MPS tensor into an intended unitary $U$ and a byproduct $V_{g}$.}
 \label{Ag}
\end{figure}

When  we measure the MPS tensor in the state $\ket{g}$, the equivariant transformation law  under $l(g)$ (Figure~\ref{equivar}) implies:
\begin{align}
\label{para}
A_g = V_g\, U = V_g\, A_{++}(\alpha_2, \alpha_1)
\end{align} 
This equation, presented graphically in Figure~\ref{Ag}, succinctly summarizes \eqref{arot} and explains its origin via the MPS symmetries. In Section~\ref{MBQCgauge}, we interpret this equation as a discrete analogue of the covariant derivative in continuum gauge theory.
 
\subsubsection{Incorporating temporal order}

\begin{figure}[h]
\centering 
\includegraphics[scale=.53]{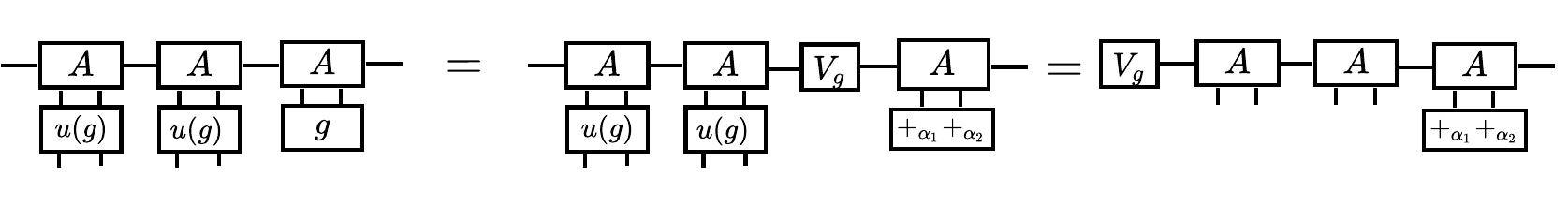}
\caption{Adaptation of measurement basis can by implemented by applying $u(g)$.}
 \label{adapts}
\end{figure}

\paragraph{Adaptation by symmetry} 
The left action $l(g)$ which defines the states $\ket{g}$ in equation \eqref{ga} accounts for the adaptation of measurement basis within a block, but does not account for the adaptation between different blocks.   To incorporate these global adaptations, we have to change the reference state in \eqref{ga} to depend on the previous outcomes. This modification can be achieved using the group action $u(g)$ \cite{SPT2}.  To explain this, let us first reformulate the basis adaptation between different blocks in terms of symmetry operations. This is illustrated in Figure~\ref{adapts}.
This first equality shows that the effect of obtaining an unwanted outcome $g$ at the first block is to produce a byproduct operator in the virtual Hilbert space, which cannot be accessed directly. However, equipped with the symmetry operation $u(g)$, our MBQC programmer can cancel out any unwanted byproduct operator $V_{g}$ in the interior of the chain by applying $u(g)$ to all yet-to-be-measured qubits.  This sends the byproduct $V_{g_{i}}$ to the end of the chain, where it joins other $V_{g}$'s in forming the byproduct $B$ in equation~(\ref{defb}). The effect of $u(g)$ on unmeasured qubits is captured by equations~(\ref{XOX}). As $u(g)$ contains only $X$ operators, future measurement outcomes are undisturbed but even/odd measurement angles are flipped. Iterating this procedure for all interior $V_{g_i}$'s reproduces equation \eqref{burelation}. 

\paragraph{Group labels for the globally adapted basis}
By iterating the adaptation procedure in Figure~\ref{adapts}, we find that the adaptation between different blocks is implemented by the cumulative action of $u(g)$ on the $j^{\rm th}$ block:
\begin{align}
u(g_{\prec j}) &= 
X^{s_{2j-3} + \ldots + s_3 + s_1} \otimes
X^{s_{2j-2} + \ldots + s_4 + s_2}\nn
    g_{\prec j} &:= g_{j-1} \ldots g_2\, g_1
\end{align}
Thus we can incorporate the global adaptation into the definition of $\ket{g}$ in by changing the reference basis to:
\begin{equation}
 u(g_{\prec j}) 
\ket{+_{\alpha_{2j}}}
\otimes 
\ket{+_{\alpha_{2j-1}}}=\ket{+_{\big((-)^{q_{2j}-s_{2j-1}}\alpha_{2j}\big)}}
\otimes 
\ket{+_{\big((-)^{q_{2j-1}}\alpha_{2j-1}\big)}}
\label{reffi}
\end{equation}
Relative to this reference state, the measurement basis at the $j$th block\footnote{Explicitly, we can write \eqref{globalbasis} as 
\begin{align}
\label{gagen} 
\ket{g_{j}}= l(g_{j})
\ket{+_{\big((-)^{q_{2j}-s_{2j-1}}\alpha_{2j}\big)}}
\otimes 
\ket{+_{\big((-)^{q_{2j-1}}\alpha_{2j-1}\big)}}, 
\quad {\rm where}~~g 
\in  \mathbb{Z}_2 \times \mathbb{Z}_2
\end{align} 
Note how the adaptive parameter in front of $\alpha_{2j}$ excludes $s_{2j-1}$, so that the action of $X_{2j}Z_{2j-1}$ in $l(g)$ reproduces equation~(\ref{orot}).
} is:
\begin{align}
\label{globalbasis} 
\ket{g_{j}}= l(g_{j}) u(g_{\prec j}) 
\ket{+_{\alpha_{2j}}}
\otimes 
\ket{+_{\alpha_{2j-1}}}
\end{align} 
Using this notation, the cluster state in the globally adapted basis is simply given by:
\begin{equation}
\cluster = 
\mathcal{N}
\sum_{g \in \mathbb{Z}_2 \times \mathbb{Z}_2} 
\Big( \langle {\rm out} | A_{g_{N}} \ldots A_{g_{2}} A_{g_{1}} | {\rm in} \rangle \Big)
|g_{N},\ldots g_{2},g_{1}\rangle
\end{equation}
Here $A_{g}$ denotes the MPS tensor measured in the globally adapted basis \eqref{globalbasis}.  To unpack this 
notation a bit, note that using the transformation laws in  Figure~\ref{equivar}, we can write:
\begin{align} \label{VA2}
    A_{g_{j}}=V_{g_{\prec i}} \left( V_{g_j} U_{f_j} \right)  V_{g_{\prec i}}^\dagger
\end{align}

\paragraph{Adaptation using stabilizers} 
Consider a single step of the `adaptation by symmetry' procedure shown in Figure~\ref{adapts}.  To cancel out the byproduct $V_{g_{i}}$ at the end of the chain, we should apply $V^{\dagger}_{g_{i}}$ to the uncontracted virtual leg of the last MPS tensor. It is now useful to view this leg as a final, $(2N+1)^{\rm st}$ qubit of a cluster chain of length $2N+1$. After doing so, the total transformation that corrects the measurement outcome $g_{i}$ is:
\begin{align} \label{Kig}
\mathcal{K}_{i}(g_i) = \overbrace{\phantom{-} V^{\dagger}_{g_i} \phantom{-}}^{(2N+1)^{\rm st}~{\rm qubit}} \left(\prod_{j=i+1}^N u_{j}(g_{i}) \right) l_{i}(g_{i})
\end{align} 
Notice that this is a product of stabilizers. In particular, for the nontrivial generators $g_i= +-$ and $g_i=-+$ we have:
\begin{align}
\label{RK}
\mathcal{K}_{i}(+-)&=Z_{2N+1}  \big( X_{2N} \cdots X_{2i+2} X_{2i} \big) Z_{2i-1}  \nn
\mathcal{K}_{i}(-+)&= X_{2N+1} \big(X_{2N-1} \cdots X_{2i+1} \big) Z_{2i}
\end{align} 
$\mathcal{K}_{i}(+-)$ is a telescoping product of $K_{j}=Z_{j+1}X_{j}Z_{j-1}$ while $\mathcal{K}_{i}(-+)$ combines $K_j$'s with the special stabilizer $X_{2N+1} Z_{2N}$, which acts at the end of the chain.  

\paragraph{Equivalent measurement records} 
A stabilizer transformation manifestly leaves $\cluster$ invariant, but conjugates the operators $O(\alpha_{i})$ as in equation~\eqref{KOK}. This in turn implies a nontrivial transformation of the classical processing data $(q_{i},s_{i})$:
\begin{align}\label{stab}
K_{j}: s_{j-1} &\to s_{j-1} +1\nn
         s_{j+1} &\to s_{j+1}+1 \\
         q_{j} &\to q_{j} +1 \nonumber
\end{align} 

\paragraph{\bartek{Relation to projective representations and extension to SPT phase}}
Observe that each stabilizer $K_j$ can be expressed in terms of projective representations of $\mathbb{Z}_2 \times \mathbb{Z}_2$. Recalling that the $j^{\rm th}$ block in the MPS description contains the $(2j-1)^{\rm st}$ and the $(2j)^{\rm th}$ qubit, we see that the stabilizers are:
\begin{align}
    K_{2j}&=r_{j+1}(+-) l_{j}(+-)\nn
    K_{2j-1}&=r_{j}(-+)l_{j-1}(-+) 
    \label{stabilizersirreps}
\end{align}
\bartek{These relations are useful because they tell us how to extend our discussion away from the cluster state---to other MBQC resource states drawn from the SPT phase. All states in a $G$-protected topologically ordered phase admit representations $l(g)$ and $r(g)$ of $G$ at a sufficiently coarse scale \cite{Schuch}. (A more direct relation between $G$-gauge theory and $G$-protected topological order is discussed in \cite{Levin2012}.) Therefore, for all SPT-ordered states the right hand sides of (\ref{stabilizersirreps}) are well defined. The assertions in this paper, which relate gauge theoretic concepts to stabilizers $K_j$, extend to other SPT states by substituting equations~(\ref{stabilizersirreps}). More details are given in Sections~\ref{sec:spt1} and \ref{ESPT}.}

\subsection{MBQC on a ring}
\label{sec:mbqcring}
In a conventional treatment, MBQC stops at equation~(\ref{chainoutput}), having extracted from a simulation of a general single-qubit gate $U$ one classical bit of computational output.  Yet common lore (technically known as the superdense coding protocol) implies that a single qubit ought to be equivalent to two bits of classical data. Working with a chain-like cluster state, we could try to recover another classical bit of data by varying the initial boundary condition $|{\rm in}\rangle \to |{\rm in}_\pm \rangle$, just like we did with the final boundary condition $\langle {\rm out}_\pm|$. Doing so would probe the other $\mathbb{Z}_2$ component of the advertised $\mathbb{Z}_2 \times \mathbb{Z}_2$ Wilson line. 

In fact, setting this up on a chain-like cluster state is an awkward exercise. The same benefit is achieved far more simply by placing the cluster state on a ring. The extra simplicity afforded by the ring topology mirrors facts from gauge theory. A closed integral of a gauge field such as \bartek{${\rm P}\!\exp\oint dx^\mu A_\mu(x)$}, called a Wilson loop, is automatically gauge invariant. Its line analogue, \bartek{${\rm P}\!\exp\int_i^f dx^\mu A_\mu(x)$}, called a Wilson line, is tainted by a gauge dependence at endpoints. For a gauge-invariant object on a line segment, a Wilson line must be combined with a charge at each end. In our discussion, the role of these charges is played by the variable boundary conditions $\langle {\rm out}_\pm|$ and $| {\rm in}_\pm\rangle$. Indeed, what achieves gauge-independence in equation~(\ref{whattomeasure}) is the transformation $\langle {\rm out}_\pm | \to \langle {\rm out}_\pm | B$, i.e. $\langle {\rm out}_\pm|$ must \bart{transform under gauge transformations}. 

\paragraph{The MBQC protocol on a ring}
With respect to the adaptive eigenbasis of $O\big((-)^{q_j}\alpha_j)$, the wavefunction for the cluster state is given by
\begin{align}\label{1} 
    \ket{1} 
= \mathcal{N}
\sum_{g_j \in \mathbb{Z}_2 \times \mathbb{Z}_2} 
{\rm Tr}( B_{g_{N}\cdots g_{1}} U ) \, \ket{g_{N}} \cdots|g_{2}\rangle |g_{1}\rangle ,
\end{align}
where $\ket{g}$ are adapted basis elements as defined in \eqref{ga}, and we have applied  equation \eqref{burelation}.   
We now observe that any $SU(2)$ matrix $U$ can be expanded in the operator basis $V_g$ as 
\begin{align}
    U = \sum_g c_g V_g 
\end{align}
The coefficients in this expansion are:
\begin{equation}
c_g = {\rm Tr} (V_g^\dagger U) / {\rm Tr} (V_g^\dagger V_g) = \tfrac{1}{2} {\rm Tr} (V_g^\dagger U)
\label{defcg}
\end{equation} 
Since $B$ takes values in $V_g = \pm V_g^\dagger$, the amplitudes ${\rm Tr} (B\,U)$ determine the coefficients $c_g$ up to a phase.

The MBQC protocol on a ring returns as computational output the probability distribution $p_g = |c_g|^2$. As anticipated, the reported data consist of two classical bits of information. They fully characterize the simulated unitary $U \in SU(2)$ as an expansion in the operator basis $V_g$. In the end, the output is sensitive only to the overall byproduct operator (\ref{defb}), which accumulates around the ring. 

In the next section, we interpret the byproduct $B$ as a Wilson loop of a $\mathbb{Z}_2 \times \mathbb{Z}_2$ gauge field or, equivalently, as a $\mathbb{Z}_2 \times \mathbb{Z}_2$ flux. In terms of individual measurement outcomes $s_j$, the computational output is the $\mathbb{Z}_2 \times \mathbb{Z}_2$-valued random variable:
\begin{equation}
o = \prod_{j} g_{j} 
\label{wloop}
\end{equation}
Once more, we emphasize that (\ref{wloop}) puts all measurement outcomes on equal footing. 

For future reference, notice that the adaptation of measurement angles on the ring can again be understood in terms of stabilizer transformations, just as it could on the chain. In particular, equations~\eqref{stab}, which generate the globally adapted basis, leave the computational output invariant. One simplification on the ring is that the corrective procedure involves only $K_j = Z_{j+1} X_j Z_{j-1}$; there are no more special stabilizers such as $X_{2N+1} Z_{2N}$.

\section{The gauge theory of MBQC}
\label{MBQCgauge}
In this section, we establish the gauge theory of MBQC. In Section~\ref{GauP}, we describe the MBQC gauge principle, first informally and then formally as a theorem. This invokes centrally a notion of `reference.' The mathematics of gauge theory knows of such an object---the section. This correspondence forms the starting point for reviewing the mathematical underpinning of gauge theory in Section~\ref{MathPin}. There we establish how the notion of gauge potential applies to MBQC. With this insight, we re-examine the classical side processing relations of MBQC in Section~\ref{ElemC}.

\bartek{The identifications between MBQC and gauge theoretic concepts are summarized in Table~\ref{summarytable}.}

\begin{table}
\centering
\begin{tabular}{r p{0.05cm} c p{0.05cm} c}
  {\bf gauge theory} & & {\bf MBQC}  & & {\bf notation / reference} \\
  \cline{1-5} \\[-0.5em]
    section & & one possible run of MBQC & & $|f_i\rangle$ in equation~(\ref{i}) \\[0.5em]
  \cline{1-1} \cline{3-3} \cline{5-5}
  \\[-0.5em]
  gauge potential & & byproduct operator & & $V_{g_i}$ in equation~(\ref{defasj}) \\[0.5em]
  \cline{1-1} \cline{3-3} \cline{5-5}
  \\[-1em]
  parallel transport & & \makecell{virtual space operator \\ induced by measurement} & & $A_{s_i}$ in equation~(\ref{defasj}) \\
    \cline{1-1} \cline{3-3} \cline{5-5}
  \\[-1em]
  gauge transformation & & \makecell{transforming between \\ computationally equivalent runs} & & \makecell{equations \\ (\ref{sgt}) and (\ref{gtf})} 
  \\[0.5em]
  \cline{1-1} \cline{3-3} \cline{5-5}
  \\[-1em]
  \makecell[cr]{section that \\ trivializes the potential} & & 
  \makecell[cc]{MBQC run already conducted \\ and the {\bf adapted} planned run} & & 
  \makecell{equations \\ (\ref{mbqcbasisgauge}) and (\ref{lastgaugefix})} \\
    \\[-1em]
    \cline{1-1} \cline{3-3} \cline{5-5}
    \\[-1em]
  & & \makecell{in physical basis: \\ property of resource state}& & 
  \makecell{$V_o U_{\rm total}$ in (\ref{genflux}) \\ is superposition~(\ref{genfluxsimple})\\ of physical fluxes} \\
  flux & &  & &  \\
  & & \makecell{in MBQC basis: \\ computational outcome} & & $V_o$ from~(\ref{defmbqcflux}) 
  \\[1em]
  \cline{1-1} \cline{3-3} \cline{5-5}
  \\[-1em]
  \makecell[cr]{measurement of flux\\ in a non-diagonal basis} & & MBQC computation & 
  & \makecell{$p_o = |\langle U^\dagger V_o| 1\rangle |^2$\\ in equation~(\ref{UV1})} \\
    \\[-1em]
 \hline
\end{tabular}
\caption{A summary of identifications between MBQC and gauge theoretic concepts.}
\label{summarytable}
\end{table}

\subsection{Gauge principle}\label{GauP}
In the circuit interpretation of MBQC, individual quantum gates can only be implemented up to a byproduct operator in the Pauli group. The byproduct operator is a priori random but known through the outcome of the measurement that implemented the gate. That is, if $U_i$ is the intended gate, then in any given run of MBQC, any of the four gates in the equivalence class
\begin{equation}\label{class}
[U_i] = \{U_i,X\, U_i, Y\, U_i,Z \,U_i \}.
\end{equation}
may be implemented, depending on the measurement outcome $s_i$ on block $i$. Thus, the procedure of gate simulation is one and the same for the entire class $[U_i]$. The in-advance unpredictable measurement result is picked by a whim of nature, and it determines the representative. 

Likewise, the person operating an MBQC  may choose a representative of $[U_i]$ as the operation {\em{intended}}. Different choices of target within the class $[U_i]$ are equivalent, because the implementation procedure can aim for the class only. Picking the intended operation is therefore a choice of reference. In the following we  denote it  as $U_{f_i} \in [U_i]$, keeping in mind that there is one choice $f_i$ per block $i$; i.e. the total gauge choice is $\textbf{f}=(f_1,..,f_N)$.

The gauge theory formulation of MBQC developed here is based on the following observation:
$$
\fbox{\parbox{11.2cm}{\textbf{MBQC gauge principle.} For any given MBQC, the reference $\textbf{f}$ can be changed without changing the quantum computation. The  changes in the reference $\textbf{f}$ are the MBQC gauge transformations.}}
$$
The principle says that an external observer of a measurement-based quantum computation who has access to the measurement record, the measurement bases, and the computational output, and who furthermore knows what the computational task is, can not infer which gauge-equivalent reference $\textbf{f}$ is used by the operator of the MBQC.

\subsubsection{Formulating the MBQC gauge principle as a theorem}

Since the framework of MBQC already exists, the MBQC gauge principle cannot just be postulated. Rather it arises as a theorem. 
We establish it here for measurement-based quantum computations on 1D cluster states, as the simplest instance. Beyond 1D cluster states, we establish it for the entire $\mathbb{Z}_2\times \mathbb{Z}_2$ SPT phase surrounding the 1D cluster state in Section~\ref{ESPT}. We conjecture the MBQC gauge principle to hold for cluster states in arbitrary dimension, and for subsystem SPT phases surrounding them.

In the review of MBQC in Section~\ref{sec:review}, and indeed in the exploration of MBQC to date, the concept of the reference $\textbf{f}$ has not been identified. Implicitly, the choice $\textbf{f}=0$ has always been assumed. This is not the only admissible choice of reference, and structural information about MBQC is revealed by realizing that this is so.

We now formulate the gauge principle at the level of the MPS tensors representing the cluster state in the MBQC measurement basis. We do this in two stages. First, we consider individual MPS matrices in isolation, which conveys the general idea. In a second step, we consider all the MPS matrices (one per block) in combination, as they are linked through the adaptation of measurement bases in MBQC. This leads us to the explicit form of the gauge transformation in Lemma~\ref{sgtExpl} below.

\paragraph{Single block level.}  Let us focus on a single MPS tensor and denote the physical measurement outcomes (two bits' worth) at the $i^{\rm th}$ block as $s_i$.  Denoting the reference for block $i$ by $f_{i}$, we consider measurements in the locally adapted basis $\ket{s_{i}}=l(g_{i})\ket{f_{i}}$ as in  eq.~\eqref{localmbqcbasis}.  Projected onto this basis, the MPS tensor decomposes as
$$
A_{s_i} = V_{g_i} U_{f_{i}},
$$
with $U_{f_{i}}:=A_{f_{i}}$ the `intended gate,' and $V_{g_i}$ the outcome-dependent Pauli byproduct. We observe that the different components of the MPS tensor $A$ of the cluster state differ only by Pauli operators. As shown in Figure~\ref{Ag}, this is a special property of the cluster tensor implied by its transformation law under the  symmetry $l(\mathbb{Z}_2 \times \mathbb{Z}_2)$. 

At this point, we realize that the splitting into an outcome-dependent Pauli part $V_{g}$ and an outcome-independent unitary $U_{f}$ is not unique.
In particular, we are not required to take $f_{i}=(+_{\alpha_{2i}}+_{\alpha_{2i-1}} )$ as the reference outcome; any outcome $f$ will do. Under a change of reference
\begin{align} \label{i}
    \quad \ket{\tilde{f}_{i}}:=l(h_{i})\ket{f_{i}}, \qquad \tilde{g}_{i} =g_{i} h^{-1} ,
\end{align}
we may write
\begin{align}\label{sgf}
\ket{s_i} =l(g_i) \ket{f_i} =l(\tilde{g_i})\ket{\tilde{f_i}},\;\; \forall i=1,..,N,
\end{align}
so the group label for $\ket{s_{i}}$ changes to $\tilde{g}_{i}$.
Each component $A_s$ of the MPS tensor $A$ may now be written as
\begin{align}
A_{s_i}=V_{g_i} U_{f_i} =V_{\tilde{g}_i} U_{\tilde{f}_i}.
\label{defasj}
\end{align} 
  For any Pauli operator $V_h\in {\cal{P}}_1$, the mapping $U_{f}\mapsto U_{\tilde{f}_i} = V_hU_{f_i}$, $V_{g_i} \mapsto  V_{\tilde{g}_i}= V_{g_i}V_h^\dagger$ keeps $A_{s_i}$ invariant; and furthermore $V_hU_{f_i}\in SU(2)$ for all $U_{f_i} \in SU(2)$, and $V_{s_i}V_h^\dagger \in {\cal{P}}_1$ for all $V_{s_i}\in {\cal{P}}_1$. The splitting of $A_{s_i}$ into an outcome-dependent Pauli part $V_{s_i}$ and an outcome-independent general unitary part $U_{f_i}$ is therefore ambiguous. This ambiguity is the origin of the MBQC gauge principle.

\paragraph{Gauge dependence at the multi-block level.} The reason we need to look at multiple MPS matrices combined is the adaptivity of local measurement basis in MBQC, according to previously obtained measurement outcomes.
This is captured by equation \eqref{gagen}, where the outcome dependence of the measurement basis at block $i$ is implemented by $u({g_{\prec i}})$. Since $u({g_{\prec i}})$ depends on the group labels $g_{j}, j \leq i$, the basis states $\ket{g_{i}}$ transform nontrivially under a change of reference on any site $j \leq i$.
Likewise the MPS tensor $A_{g_{i}}$ measured in the adapted basis depends on reference choices on sites $j \leq i$.
To make explicit the dependence of MBQC on change of references, let us introduce  reference labels on the globally adapted basis states and the corresponding MPS tensors : 
\begin{align}\label{ref}
    \ket{g_{i} } &:= \ket{g_{i}, f_{i}} =  l(g_{i}) u({g_{\prec i}}) \ket{f_{i}} \nn
    A_{g_{i}} &:= A_{g_{i},f_{i}} =  V_{g_{\prec i}} \left( V_{g_i} U_{f_i} \right)  V_{g_{\prec i}}^\dagger
\end{align}
Similarly, classical side processing data depend on the reference via their explicit dependence on group labels: 
\begin{align} 
 (q_{2i}, q_{2i-1}) &= g_{\prec i} 
 \qquad \textrm{(as element of $\mathbb{Z}_2 \times \mathbb{Z}_2$)} \\
\label{cprGTb}
o &= \prod_{i} g_{i} 
\end{align}
If we define  
\begin{align}\label{gf1}
A_{\mathbf{g},\mathbf{f}} = \prod_{i} A_{g_{i},f_{i}} ,\quad V_{o} = \prod_{i} V_{g_{i} } ,\quad  U_{\mathbf{f}}=  \prod_{i} U_{f_{i}},
\end{align}
the total MPS wavefunction for the cluster chain is given by
\begin{align}\label{gf}
  A_{\mathbf{g},\mathbf{f}} =  V_{o}   U_{\mathbf{f}}
\end{align}
where we have used byproduct propagation to obtain the RHS as in equation \eqref{burelation}.

Equations~(\ref{ref}-\ref{gf}) summarize all the relevant formulas of MBQC. 
Inspection of these relations motivates a further constraint to impose on the change of reference. From the circuit model perspective, $U_\textbf{f}$ is the unitary that MBQC realizes and to which the classical output of MBQC refers; i.e. the MBQC output reveals properties of $U_\textbf{f}$. The change of reference should not change that interpretation. Thus, while the locally implemented pieces $U_{f_i}$ may change under the gauge transformation, their cumulative product $U_\textbf{f}$ should not.  

We thus arrive at the following definition for the notion of MBQC gauge transformations.
\begin{Def}\label{GTdef}
An MBQC gauge transformation is a change of reference $\ket{\textbf{f} }\mapsto \ket{ \tilde{\textbf{f}}}$ and accompanying change of the group labels for the byproducts $\textbf{g}=(g_1,..,g_N)$ such that (i) all MPS matrices $A_{g_i,f_i}$ are preserved, up to possible global phases, and (ii) the total MBQC-implemented unitary is preserved,
\begin{equation}\label{SGT}
U_{\textbf{f}} \mapsto U_{\tilde{\textbf{f}}} = U_{\textbf{f}}.
\end{equation}
\end{Def}
\noindent In a different terminology, the gauge transformations defined above are called 
proper gauge transformations whereas changes of reference that satisfy condition (i) but do not satisfy (ii) are called `large' gauge transformations. 

\bartek{In the language of conventional gauge theory, condition~(\ref{SGT}) can be viewed as a consequence of Gauss's law: you cannot create a flux by a gauge transformation. In the context of the cluster state $\cluster$, (\ref{SGT}) follows most directly from the fact that the resource state is invariant under operators $K_j$ defined in (\ref{defstab}). Both perspectives are mutually consistent: in a quantum gauge field theory Gauss's laws (as constraints) act as gauge transformations whereas on the cluster state gauge transformations are generated by $K_j$. (This fact can be seen by comparing equations~(\ref{stabilizersirreps}) and (\ref{gtf}) below.)}

Eq.~(\ref{gf}), in combination with Definition~\ref{GTdef}, leads to the following observation:
\begin{Lemma}\label{sgtInvar}
The cumulative MPS matrix $A_{\textbf{g},\textbf{f}}$, the target unitary $U_\textbf{f}$ and the cumulative byproduct $V_o$ are, up to possible global phases, invariant under the  MBQC gauge transformations.
\end{Lemma}
\noindent
{\em{Proof of Lemma~\ref{sgtInvar}.}} $A_{\textbf{g},\textbf{f}}$ and $U_\textbf{f}$ are invariant by Definition~\ref{GTdef}. Therefore, by Eq.~(\ref{gf}), the cumulative byproduct $V_o$ is also invariant. $\Box$ 
\medskip

We now provide an explicit construction of the gauge transformations according to Definition~\ref{GTdef}. We have the following result:
\begin{Lemma}\label{sgtExpl}For all $i=1,..,N-1$ and all $h\in \mathbb{Z}_2\times \mathbb{Z}_2$, the transformations $h^{(i)}$ defined by
\begin{equation}\label{sgt}
\begin{array}{rclcrcl}
U_{f_i} & \mapsto & U_{\tilde{f}_i} =V_h U_{f_i},&& V_{g_i} &\mapsto& V_{\tilde{g}_i} = V_{g_i}V_h^\dagger,\\
U_{f_{i+1}} & \mapsto & U_{\tilde{f}_{i+1}} =U_{f_{i+1}} V_h^\dagger,&&  V_{g_{i+1}} &\mapsto& V_{\tilde{g}_{i+1}} = V_{g_{i+1}}V_h.
\end{array}
\end{equation}
are MBQC gauge transformations.
\end{Lemma}
 The gauge transformations Eq.~(\ref{sgt}), together with the equivariant transformations $l(\cdot)$ and $r(\cdot)$ of the cluster state MPS tensor displayed in Figure~\ref{equivar}, imply the following transformation on the reference $\ket{\textbf{f}}$:
\begin{equation}\label{gtf}
h^{(i)}: \;\; \ket{f_i}\mapsto l( h) \ket{f_i},\; \ket{f_{i+1}} \mapsto r(h) \ket{f_{i+1}},
\end{equation}
with all other components of $\textbf{f}$ unchanged.  The corresponding  change in group labels is:
\begin{align}\label{gtg}
 \tilde{g}_{i} = g_{i}h^{-1} ,\qquad \tilde{g}_{i+1} =h g_{i+1}.
\end{align}

\noindent
{\em{Proof of Lemma~\ref{sgtExpl}.}} We have to check that the transformations according to eq.~(\ref{sgt}) satisfy Properties (i) and (ii) of Definition~\ref{GTdef}. 

Property (i). The gauge transformation (\ref{sgt}), for any fixed $i$, affects non-trivially only blocks $i$ and $i+1$. Checking the corresponding MPS matrices, we find:
\begin{equation}
\begin{array}{rcl}
A_{g_i,f_i} &=& V_{g_{\prec i}} \left( V_{g_i} U_{f_i} \right)  V_{g_{\prec i}}^\dagger\\ 
&\mapsto& V_{\tilde{g}_{\prec i}} \left( V_{\tilde{g}_i} U_{\tilde{f}_i} \right)  V_{\tilde{g}_{\prec i}}^\dagger\\ 
&\propto&V_{g_{\prec i}} \left( V_{g_i} V_h^\dagger V_h U_{f_i} \right)  V_{g_{\prec i}}^\dagger\\ 
&=&V_{g_{\prec i}} \left( V_{g_i} U_{f_i} \right)  V_{g_{\prec i}}^\dagger\\
&=& A_{g_i,f_i}, \label{agf1}
\end{array}
\end{equation}
and
\begin{equation}
\begin{array}{rcl}
A_{g_{i+1},f_{i+1}} &=& V_{g_{\prec i+1}} \left( V_{g_{i+1}} U_{f_{i+1}} \right)  V_{g_{\prec {i+1}}}^\dagger\\ 
&\mapsto & V_{\tilde{g}_{\prec i+1}} \left( V_{\tilde{g}_{i+1}} U_{\tilde{f}_{i+1}} \right)  V_{\tilde{g}_{\prec {i+1}}}^\dagger\\ 
&\propto & V_{g_{\prec i+1}} V_h^\dagger  \left( V_{g_{i+1}} V_h U_{f_{i+1}} V_h^\dagger \right)  V_h V_{g_{\prec {i+1}}}^\dagger\\ 
&\propto &  V_{g_{\prec i+1}} \left( V_{g_{i+1}} U_{f_{i+1}} \right)  V_{g_{\prec {i+1}}}^\dagger\\ 
&=& A_{g_{i+1},f_{i+1}}. \label{agf2}
\end{array}
\end{equation}
Thus, up to possible global phases, the MPS matrices are invariant. Property (i) is satisfied.\smallskip

Property (ii). Using~(\ref{sgt}), we have:
$$
\begin{array}{rcl}
U_\textbf{f} &=& U_{f_N}..U_{f_{i+2}}U_{f_{i+1}}U_{f_{i}}U_{f_{i-1}}..U_{f_1}\\
&\mapsto & U_{f_N}..U_{f_{i+2}}\left(U_{f_{i+1}}V_h^\dagger\right)\left(V_h U_{f_{i}}\right) U_{f_{i-1}}..U_{f_1}\\
&=& U_{f_N}..U_{f_{i+2}}U_{f_{i+1}}U_{f_{i}}U_{f_{i-1}}..U_{f_1}\\
&=& U_\textbf{f}.
\end{array}
$$
Thus, Property (ii) holds as well. $\Box$\medskip

We can now state the MBQC gauge principle as a theorem:
\begin{Theorem}[MBQC gauge principle]\label{gtT}
For any given MBQC on 1D cluster states, the measurement bases $\{ \ket{g_{i },f_{i }}\} $ chosen as a function of prior measurement record ${g_{\prec i}}$, the output $o$ given the measurement record $\textbf{g}$, and the probability distribution of the computational output $o$, $p(o)= |\text{Tr}(U_\textbf{f} V_o)|^2$, are invariant under gauge transformations~(\ref{sgt}-\ref{gtg}). 
\end{Theorem}
\noindent
This has the following consequence: An external observer, who tracks an MBQC computation and has access to all measurement settings, the measurement record and the computational output, cannot infer which reference $\textbf{f}$ is used in the internal classical side processing.\medskip

\noindent
{\em{Proof of Theorem~\ref{gtT}.}} (i) The local measurement bases: Under the gauge transformation $h^{(i)}$, the local measurement basis $\{|g_{i+1},f_{i+1}\rangle, g_{i+1} \in \mathbb{Z}_2\times \mathbb{Z}_2\}$ at block $i+1$ transforms as:
\begin{align}
 l(g_{i+1})   u({g_{\prec i+1}})\ket{f_{i+1}} &\to  l(\tilde{g}_{i+1}) u({\tilde{g}_{\prec i+1}})  \ket{\tilde{f}_{i+1}}\nn
  &= l(h g_{i+1}) u({g_{\prec i+1}}) u(h) r(h)  \ket{f_{i+1}}\nn
  &= l(g_{i+1}h) u({g_{\prec i+1}}) l(h)  \ket{f_{i+1}}\nn
  &\propto l(g_{i+1}h)l(h) u({g_{\prec i+1}} ) \ket{f_{i+1}}\nn
  &=  l( g_{i+1}) u({g_{\prec i+1}} ) \ket{f_{i+1}}
\end{align}
In the first line, we have used the gauge transformations (\ref{gtf}-\ref{gtg}). In the third line we have used $u(h)=l(h)r(h)$, cf. eq.~(\ref{LinRep}), and in the fourth we commuted $u({g_{\prec i+1}})$, and $ l(h)$ at the cost of an irrelevant sign. We thus find that the measurement basis for block $i+1$ is indeed invariant under the gauge transformations $h^{(i)}$.

On the $i^{\rm th}$ block,  we have:
\begin{align}
    l(g_{i}) u(g_{\prec i}) \ket{f_{i}} &
    \to  l(\tilde{g}_{i}) u({\tilde{g}_{\prec i}}) \ket{\tilde{f}_{i}}\nn
    &= l(g_{i} h) u(g_{\prec i})  l(h) \ket{f_{i}} \nn
    &\propto  l(g_{i}h)  l(h)  u(g_{\prec i})  \ket{f_{i}} \nn
    &\propto  l(g_{i} )  u(g_{\prec i})  \ket{f_{i}}
\end{align}
On all other blocks, the gauge transformation $h^{(i)}$ acts trivially. Thus, all local measurement bases are gauge-invariant.

{(ii) Single-shot computational output $o$:} A gauge transformation $h^{(i)}$ transforms the computational output as follows:
$$
\begin{array}{rcl}
o &=& g_1.. g_N\\
&\mapsto& g_1..g_{i-1}(g_i h^{-1})(h g_{i+1})g_{i+2}..g_N\\
&=& g_1..g_N\\
&=& o.
\end{array}
$$
Thus, the output $o$ is invariant under $h^{(i)}$, for any $h\in \mathbb{Z}_2\times \mathbb{Z}_2$, for any $i$.

{(iii) Output statistics:} Since $o$ is gauge invariant per item (ii), so is $V_o$, up to a possible phase. $U_\textbf{f}$ is invariant by assumption; cf. eq.~(\ref{SGT}). Therefore $p(o)=|\text{Tr}(U_\textbf{f} V_o)|^2$ is gauge-invariant. $\Box$

\subsection{Mathematical underpinning of gauge theory}
\label{MathPin}
We now formalize the objects encountered in Section~\ref{GauP}. Our first goal is to review the concept of `section,' and, for the application to MBQC, identify it with the reference $\textbf{f}$.
 \smallskip

\subsubsection{Fiber bundles, sections, and gauge transformations}\label{FBS}
Gauge theory is a theory of fiber bundles.\footnote{\bartek{By `gauge theory', we mean a quantum system whose states describe quantized holonomies of a fiber bundle with a fluctuating connection. The specific gauge theory we describe here belongs to the class of Dijkgraaf-Witten gauge theories defined in \cite{dw}.}} Accordingly, our next step is to present a (discrete) fiber bundle, which describes the geometry of an MBQC computation. We begin with a lightning review of fiber bundles. 

A fiber bundle is a family of identical spaces (fibers) $F_b$, which are labeled by elements of another topological space $\mathcal{B} \ni b$, called the base space. The union of all fibers is called `total space' and denoted $\mathcal{E}$:
\begin{equation}
    \mathcal{E} = \bigcup_{b \in \mathcal{B}} F_b
\end{equation}
Locally, any fiber bundle looks like the direct product $F \times \mathcal{B}$. Globally, however, the direct product structure breaks down because the different fibers can be arranged over $\mathcal{B}$ in a nontrivial way. Common examples of nontrivial bundles include the M{\"o}bius strip, in which the interval $[0,1] =: F_b$ gets reflected as $b$ goes around the circle $S^1 =: \mathcal{B}$, or the tangent spaces $T_b =: F_b$ over a manifold $\mathcal{M} =: \mathcal{B}$ if the manifold is curved.

It is useful to distinguish two types of bundles. First, a principal fiber bundle $\mathcal{G}$, also called a gauge bundle, has as fibers copies of a symmetry group $G$. Given a gauge bundle and a choice of representation $R$, we can define an associated vector bundle $\mathcal{V}_{R}$ whose fibers are the vector spaces on which the representation $R$ acts.  
In the conventional setting of continuum gauge theory, $\mathcal{V}_{R}$ corresponds to a choice of matter field transforming in the representation $R$.  The gauge field then couples to this matter field via covariant derivatives in the chosen representation.

\paragraph{The MBQC gauge bundle} 
The structure of the MBQC gauge bundle is manifest in the MPS description of the resource state.   The base space $\mathcal{B}$ is the one-dimensional lattice where the physical qubits live. In the cluster state, due to the symmetries of the MPS tensor, we will identify a single point $j \in \mathcal{B}$ with a block of two physical qubits.  

The fiber above each block is the group $\mathbb{Z}_2 \times \mathbb{Z}_2$, which permutes the four possible measurement outcomes. Equivalently, each fiber of the MBQC gauge bundle consists of the measurement outcomes themselves. We anticipated this in equation~(\ref{ga}) when we labeled measurement outcomes with group elements. 

As emphasized above that equation, a labeling of a fiber with group elements is not unique since we can always act with the group on itself and get a new labeling.  To coordinatize a fiber with group elements, we must therefore choose a so-called section\footnote{In a general fiber bundle there may not exist a global section, but this subtlety does not arise in this paper.}: an outcome in every fiber, which is labeled with the identity element of the group. We did that in equation~(\ref{ga}) by choosing the outcome $(s_{2j}, s_{2j-1}) = (0,0)$---equivalently, the basis vector $|+_{\alpha_{2j}}\rangle \otimes |+_{\alpha_{2j-1}}\rangle$---as a reference. Note that, from the MBQC programmer's point of view, the identification of the state $|+_{\alpha_{2j}}\rangle \otimes |+_{\alpha_{2j-1}}\rangle$ with the group element $++ \in \mathbb{Z}_2 \times \mathbb{Z}_2$ is an arbitrary choice. Other measurement outcomes can be identified with the identity element equally well. To maintain the visual correspondence between measured states $|\pm_{(\cdots)}\rangle$ and group elements, the programmer can always modify the reference measurement angles, for example:
\begin{equation}
|-_{(-\alpha_2)}\rangle \otimes |-_{\alpha_1}\rangle = |+_{(\pi-\alpha_2)}\rangle \otimes |+_{(\pi + \alpha_1)}\rangle
\end{equation}

More formally, a section is one choice of element in each fiber: 
\begin{equation}
f\!:~ \mathcal{B} \to \mathcal{E} \qquad \textrm{such that}~~j \to f_{j} \in F_{j} 
\label{gaugechangef}
\end{equation} 
In MBQC, we identify the set of reference outcomes $\bold f $ with a section $f$ of the gauge bundle.

\begin{figure}[t]
\centering 
\includegraphics[scale=.4]{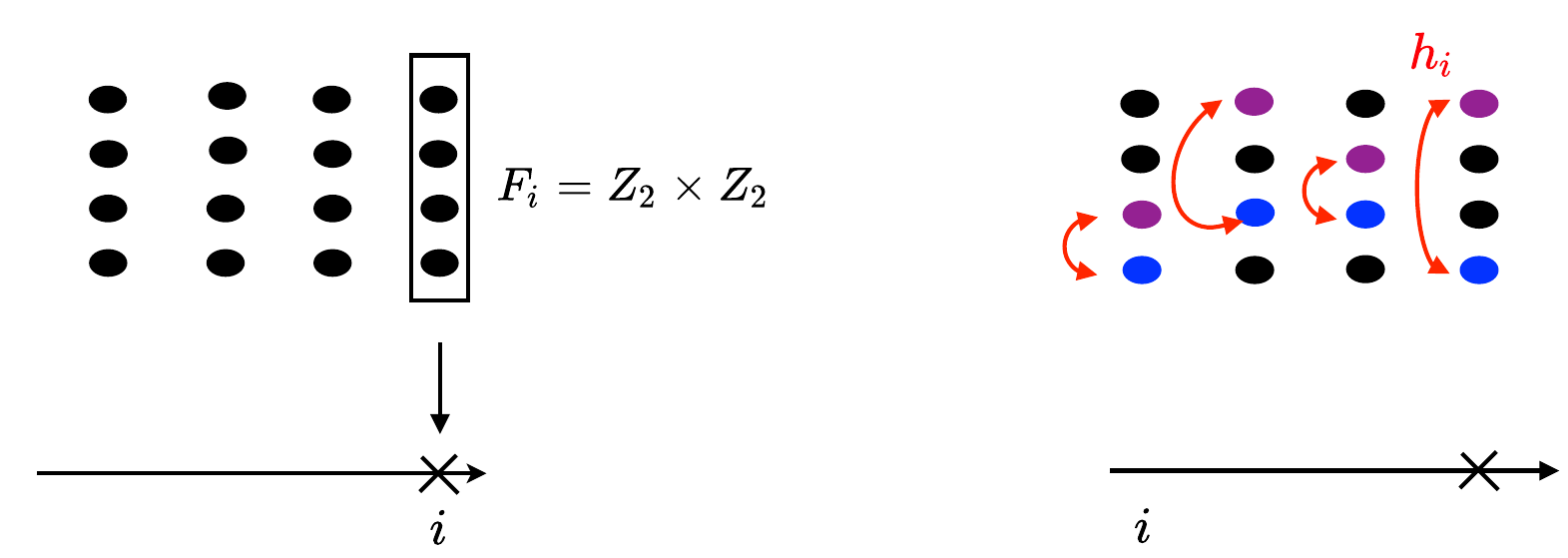}
\caption{The MBQC gauge  bundle is defined by the action of the $\mathbb{Z}_2 \times \mathbb{Z}_2 $ transformations $l_{i}(g)$ on the measurement outcomes. The orbits form the fiber $\{ f_{i} \} $ over a block, which is just a copy of $Z_{2}\times Z_{2}$.    The purple and blue dots describe two sections of this bundle, corresponding to two sets of measurement outcomes.   The sections are related by a gauge transformation, denoted by $h_{i}$.}
 \label{Z2Z2bundle}
\end{figure}

A gauge transformation $h$ is a map that relates two sections $f_j$ and $\tilde{f}_j$:
\begin{equation}\label{hgauge}
h\!:~ \mathcal{B} \to G \qquad \textrm{such that}~~j \to h_j \quad {\rm with} \quad \tilde{f}_{j} = h_j f_j
\end{equation}
This is depicted in the right of Figure~\ref{Z2Z2bundle}. 

Note the subtle difference between a section and a gauge transformation. The former maps $\mathcal{B}$ to the fibers while the latter maps $\mathcal{B}$ directly to the group. This is because $h_{j}=\tilde{f}_{j}{f}^{-1}_{j}$ is a relative construct, from which the ambiguity in labeling fibers cancels out. 

\paragraph{The associated bundle} 
The associated bundle $\mathcal{V}_{R}$ for MBQC has the same base space as the gauge bundle, but the fibers are replaced by the virtual Hilbert space, which transforms under the projective representation of $\mathbb{Z}_2 \times \mathbb{Z}_2$.   More specifically, we define $F_{j}$ on the $j^{\rm th}$ block to be the Hilbert space on the left virtual edge of the MPS tensor.   A gauge transformation $h_{j}$ acts on the associated bundle via the projective representation $V_{h_{j}}$, which rotates each fiber of $\mathcal{V}_{R}$.  The virtual qubit in the circuit simulation thus plays the role of the `matter field' in the analogy with ordinary field theory.     

\begin{figure}[h!]
\centering 
\includegraphics[scale=.15]{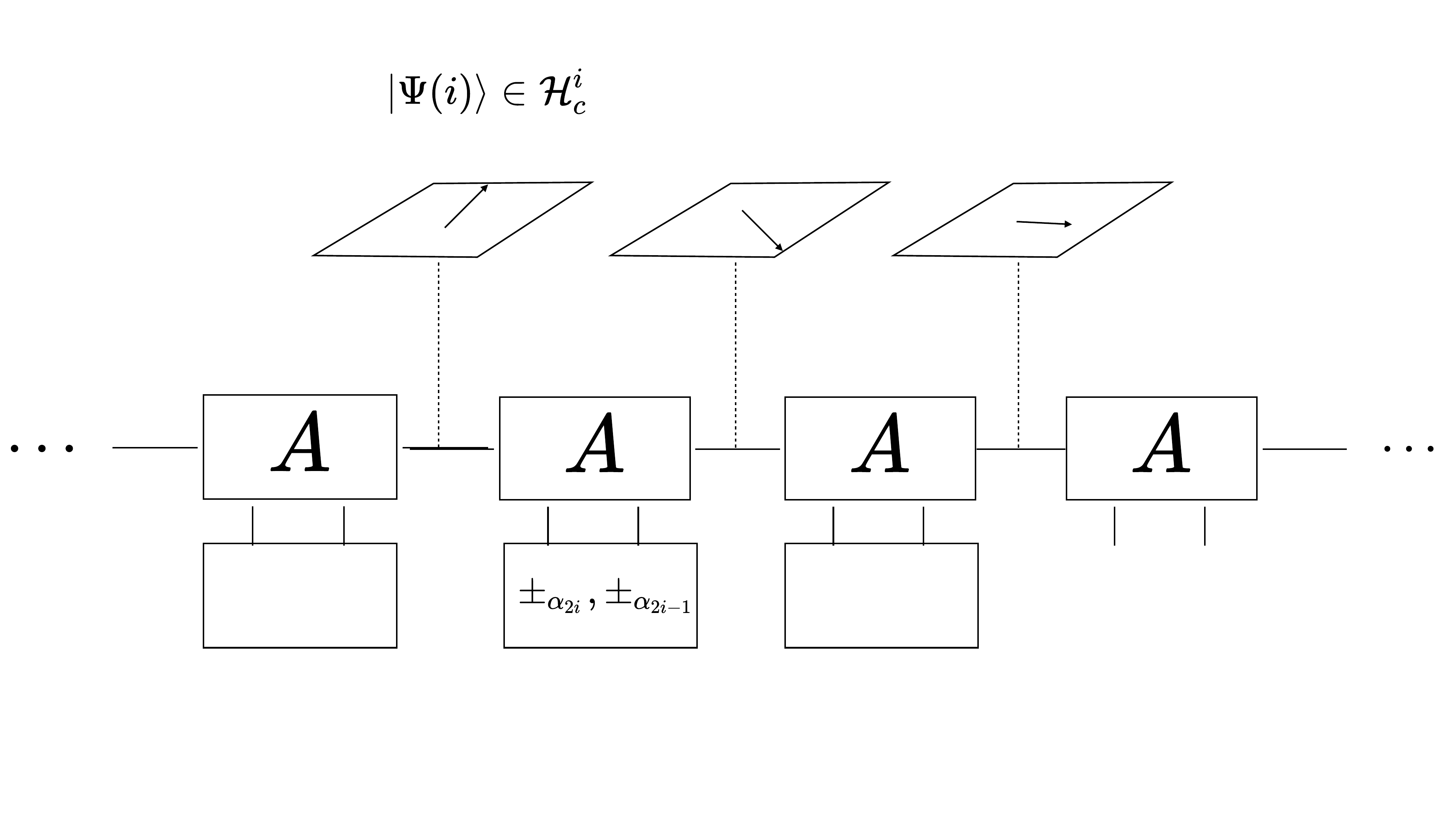}
\caption{The fibers $F_{i}=\mathcal{H}^{i}_{c}$ of the vector bundle $\mathcal{V}$ are copies of the virtual Hilbert space where the virtual qubit lives.} 
\label{V}
\end{figure}

\subsubsection{Parallel transport}
\label{PT}
\label{sec:transport}
A key concept in the study of fiber bundles is parallel transport. It sets a convention for what it means for a trajectory in the total space $\mathcal{E}$ to travel entirely in a `horizontal direction,' with no component of motion in the fiber. (In this nomenclature, motion over one point $j$ in the base space---that is entirely within $F_j$---is `vertical.') Without a definition of parallel transport there is no canonical way of comparing fibers over different base points.  Parallel transport relates different fibers to one another in a definite, albeit non-canonical, way.  It is specified by a linear map between adjacent fibers $F_{j-1}$ and $F_{j}$.
In MBQC, parallel transport is generated by the MPS tensor $ A_{s_{j}} = A_{s_{j}} \big((-)^{q_{2j}} \alpha_{2j}, (-)^{q_{2j-1}}\alpha_{2j-1} \big) $ in the adapted basis:
\begin{align}\label{AF}
A_{s_{j}}: F_{j-1} \to F_{j},
\end{align} 

We thus identify parallel transport on the bundle $\mathcal{V}_{R}$ with the quantum teleportation of the virtual qubit. In MBQC, parallel transport is induced by measurements. Prior to a measurement, the definition of parallel transport is fluctuating; we called this circumstance a fluctuating circuit in Section~\ref{cluster}. Measurements in the MBQC basis force parallel transport to take on a definite, classical form.

\paragraph{Gauge connection } 
The generator of local parallel transport $A_{s}$ is a gauge invariant linear map describing a horizontal motion between neighboring fibers. It is independent of the group label assigned to the measurement outcome $s$.  On the other hand, the gauge connection describes a gauge dependent motion along $F_{i}$ that depends on a choice of section.  To determine the MBQC gauge connection, recall that a choice of section provides a group label for the measurement outcome $s$ according to 
\begin{align}
s =gf=\tilde{g}\tilde{f} 
\end{align}  
We think of $s$ as an element of the fiber $F_j$ in the gauge bundle, which has coordinates $g$ and $\tilde{g}$ relative to two reference sections $f$ and $\tilde{f}$. Using the relation \eqref{para},  we can write the parallel transport as
\begin{align}
\label{VA}
A_{s}=V_{g} U_{f} =V_{\tilde{g}} U_{\tilde{f}},
\end{align} 
where we used the notation $U_{f}=A_{f}$ to emphasize that $A_{f}$ is the (desired) local circuit simulation. 

Equation~\eqref{VA} shows that a gauge choice decomposes the parallel transport operator into two parts. The first is a `na{\"\i}ve motion' $U_f$, which can contain both horizontal and vertical motion. In the total space $\mathcal{E}$, this na{\"\i}ve motion simply follows the section $f$. The second part, $V_g$, is the gauge connection. It is a correction term, whose action is entirely vertical---that is, it maps the fiber to itself. Its job is ensure that the combination $V_g U_f$ generates parallel transport.   Note that, when viewed in isolation, both $V_{g}$ and $U_{f}$ are ambiguous up to multiplication by $V_{h}$ for arbitrary $h$ in the gauge group; only their product has invariant meaning. \bart{These are MBQC gauge transformations:} the freedom to reset the section $f$ and the corresponding ambiguity in the byproduct $V_g$.

\paragraph{Parallel transport in the globally adapted basis}
The MPS tensor $A_{s}$ in equation \eqref{VA} describes the evolution a virtual qubit $\ket{\psi}$ in the locally adapted basis defined in \eqref{localmbqcbasis}.  Measurements in that basis produce the virtual evolution
\begin{align}
    \ket{\psi}  \to  \cdots V_{g_i}U_{f_{i}}\cdots V_{g_{1}} U_{f_{1}} \ket{\psi} 
\end{align}
However, MBQC involves measurement in a globally adapted basis, in which the MPS tensor $A_{s_{i}}$ is conjugated by the byproduct operator $V_{g\prec i}$ as in \eqref{VA2}. Na{\"\i}vely, the virtual evolution is more complicated, but if propagate the byproducts as described in Section~\ref{sec:review}, we again get a simple form of the virtual evolution: 
\begin{align} 
   \ket{\psi}  \to  \big(\cdots V_{g_{2}} V_{g_{1}}\big)\big( \cdots U_{f_{2}} U_{f_{1}} \big)  \ket{\psi} 
   \label{virtualev}
\end{align}
Thus, after the $i^{\rm th}$ measurement, we can interpret $ \big(\prod_{j \leq i}  V_{j}\big) \big(\prod_{j \leq i} U_{f_{j}} \big)  $ as total parallel transport, expressed as a combination of the `na{\"\i}ve' motion $\big(\prod_{j \leq i} U_{f_{j}} \big)$ and a correction 
$\big(\prod_{j \leq i}  V_{j}\big)$.

\paragraph{Gauge connection on the principal bundle}
We can also define a gauge connection directly on the gauge bundle $\mathcal{G}$. 
Given a history of measurement outcomes 
\begin{align}
s_{j} =g_{j} f_{j}\qquad j=1,\cdots N
\end{align} 
and a gauge choice $f$, we define the connection on the gauge bundle to be $\textbf{g}=(g_{1},\cdots,g_{N})$. Thus, the measurement record, labelled by a set of elements of $\mathbb{Z}_2 \times \mathbb{Z}_2$, is the MBQC gauge field! This is because the connection on the vector bundle is $V_{g_j}$ and the two bundles are related to one another by the representation map. 
 Furthermore, we can define the gauge potential at block $j$ as the pair of binary numbers $\vec{a}_{j}=(a_{j:1},a_{j:2})$ satisfying:
\begin{align}
    g_{j}=\big( (-1)^{a_{j:1}}  , (-1)^{a_{j:2}}\big)
\end{align}
Thus, the gauge potential is just a $\mathbb{Z}_2 \times \mathbb{Z}_2$ variable like $g_{j}$, written in additive notation.

\subsection{Holonomies and fluxes} 
\label{hf}
\begin{figure}[h]
    \centering
  \includegraphics[scale=.25]{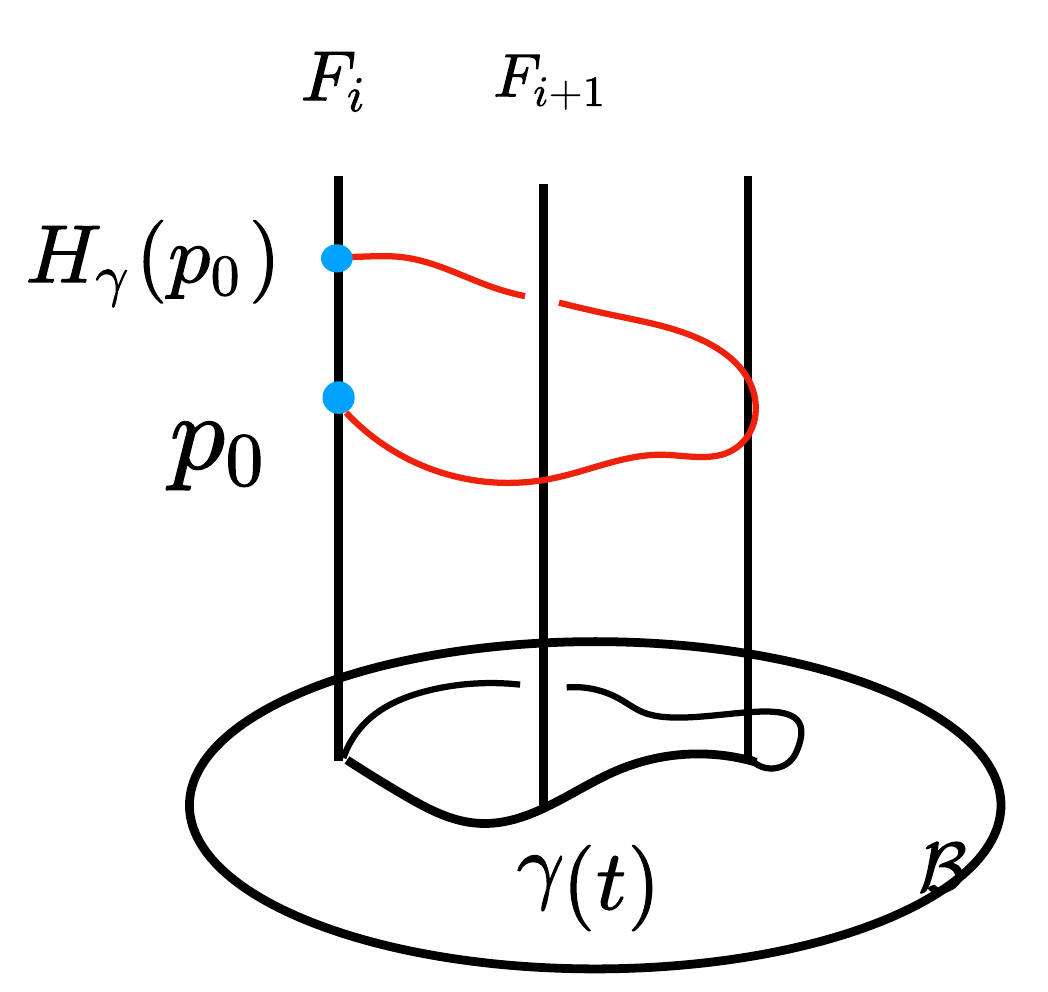}
\caption{A holonomy $H_{\gamma}$ on the \textit{principal} fiber bundle is defined by lifting a curve $\gamma(t)$ from the base space to the total space.  Given a section, this curve is a choice of group element $g_{i}$  at each fiber describing local parallel transport, which combines into a holonomy $o =\prod_{i} g_{i} $.} 
\label{holonomy}
\end{figure}

Observables in gauge theory are fluxes through closed loops. In mathematical terms, a (classical) flux corresponds to a property of a fiber bundle called holonomy. It is computed by the product of parallel transport generators around a given loop:
\begin{equation}
{\rm flux} \leftrightarrow {\rm holonomy} = \prod_{\rm loop} \textrm{(parallel transport)}
\label{fluxwords}
\end{equation}
This formula is illustrated in Figure~\ref{holonomy}. As reviewed in the previous subsection, parallel transport is a map from one fiber to another: $F_{j} \to F_{j+1}$. A composition of parallel transport around a closed loop therefore gives a transformation of the initial fiber to itself:  
\begin{equation}
{\rm holonomy}: F_1 \to F_1
\label{eqholonomy}
\end{equation}
Geometrically, this transformation tells us whether (and how) a fiber bundle is `twisted.' In the present context, where fibers are labeled by group elements, we expect four distinct holonomies that send $|g\rangle \in F_1$ to $|hg\rangle \in F_1$, where $h \in \mathbb{Z}_2 \times \mathbb{Z}_2$. Physically, the `twisting' of the fiber bundle is interpreted as a classical value of flux. 

\paragraph{Physical flux} Substituting the definition of parallel transport in~(\ref{AF}) and equation~(\ref{virtualev}), we get:
\begin{equation}
{\rm flux} = \prod_{j = 1}^N A_{g_j, f_j} = \left(\prod_{j=1}^N V_{g_j}\right) \left(\prod_{j=1}^N U_{f_j}\right)
= V_o U_{\rm total}
\label{genflux}
\end{equation}
This is the flux after a complete set of MBQC measurements, which force the fluctuating parallel transport operators to take on define values $A_{g_j, f_j}$. 

Equation~(\ref{genflux}) does not correspond to any one element $h$ of $\mathbb{Z}_2 \times \mathbb{Z}_2$, so how to interpret this equation? We view it as a quantum mechanical superposition of fluxes. To do so, expand
\begin{equation}
U_{\rm total} = \sum_{g \in \mathbb{Z}_2 \times \mathbb{Z}_2} c_g V_g
\label{udecomp}
\end{equation}
in the $V_g$, which form a complete basis in the space of $SU(2)$ matrices. Then equation~(\ref{genflux}) becomes:
\begin{equation}
{\rm flux} = \sum_{g \in \mathbb{Z}_2 \times \mathbb{Z}_2} c_{g}\, b(o,g) V_{o \cdot g},
\label{genfluxsimple}
\end{equation}
 where the projective phases $b(o,g)$ are defined by $V_o V_g = b(o,g) V_{o \cdot g}$ (These are never observed in MBQC). The right hand side is a linear combination of the four classically allowed holonomies $o \cdot g \in \mathbb{Z}_2 \times \mathbb{Z}_2$. 

Since classical fluxes correspond to differently twisted bundles, states of a quantum gauge theory naturally describe superpositions of distinct bundles.
\begin{figure}[h]
    \centering
  \includegraphics[scale=.45]{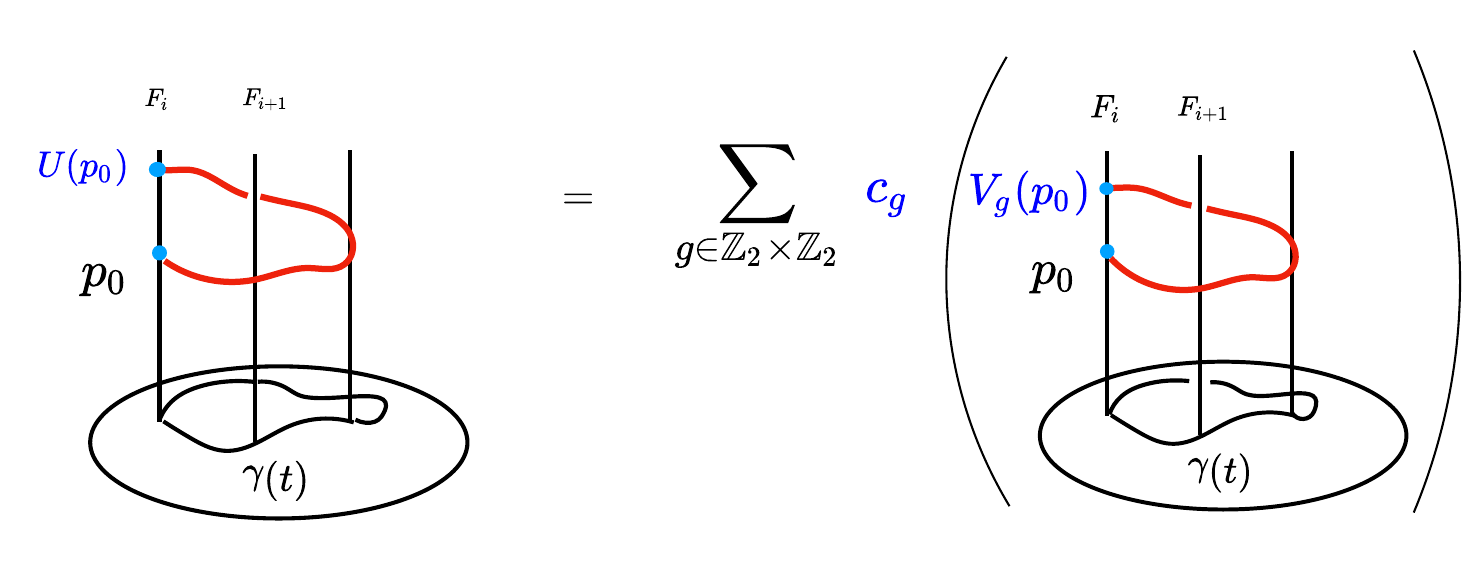}
\caption{This figure illustrates the superposition of vector bundles $\mathcal{V}_{g}$ with holonomy $V_{g}$.  In contrast with the gauge bundle where the fibers are copies of the gauge group, the fibers of $\mathcal{V}_{g}$ are  virtual Hilbert spaces $\mathbb{C}^{2}$ where the virtual qubit lives.  On the right, each term in the sum describes the parallel transport of an element $p_{0}\in \mathbb{C}^2$ around a closed loop, which transforms $p_{0}$ by a holonomy matrix $V_{g}$.   MBQC superposes such holonomies to obtain a general $SU(2)$ transformation $U$, which gives the simulated unitary. } 
\label{sbundle}
\end{figure}
A special case of (\ref{genfluxsimple}) occurs in an MBQC simulation of $U_{\rm total} = 1$ or a Pauli matrix. In this circumstance the superposition~(\ref{genfluxsimple}) involves a single term with $o \cdot g = ++ \in \mathbb{Z}_2 \times \mathbb{Z}_2$, so the outcome $o$ of MBQC is deterministic. We can understand such deterministic MBQC as measuring the trivial flux of the initial resource state $|1\rangle$. Non-deterministic MBQC, in contrast, projects the resource state onto superpositions of fluxes, which are defined in equation~(\ref{genfluxsimple}). We examine those states, and explain MBQC as a flux measurement in a non-diagonal basis, in Section~\ref{sec:quantumgauge}.

\paragraph{MBQC flux}
Consider one MBQC run, during which the resource state gets projected onto a superposition of fluxes defined in equation~(\ref{genfluxsimple}). The MBQC technician records the computational outcome 
\begin{equation}
o = g_N \ldots\, g_2\, g_1 \in \mathbb{Z}_2 \times \mathbb{Z}_2
\label{defmbqcflux}
\end{equation}
where $V_o = B$ from equation~(\ref{burelation}). From the technician's perspective, it is tedious and pedantic to say that she has projected the resource state onto a superposition of flux states; identifying flux with $o$ directly is equally informative but far simpler. 

In the remainder of the paper, we refer to $o \in \mathbb{Z}_2 \times \mathbb{Z}_2$ as `MBQC flux.' Notice that MBQC fluxes could be made consistent with equation~(\ref{fluxwords}) if we identified `MBQC parallel transport' with the individual measurement outcomes $g_j \in \mathbb{Z}_2 \times \mathbb{Z}_2$:
\begin{equation}
\textrm{`MBQC parallel transport'} = g_j~\textrm{measured on the $j^{\rm th}$ block}
\end{equation}
As we presently explain, this definition depends on subjective intentions of the MBQC technician. This is why it is useful to distinguish MBQC flux and `MBQC parallel transport' from their physical counterparts: only the latter are free of teleological considerations.

\paragraph{MBQC flux versus physical flux}
Consider a run of deterministic MBQC, in which every qubit is measured in the $X$-basis. Two quantum programmers, Alice and Bob, observe the run and jot down all measurement outcomes. Whereas Alice intends to use this MBQC run as a simulation of $U_{\rm total}^{\rm Alice} = 1 = V_{++}$, Bob intends to simulate $U_{\rm total}^{\rm Bob} = X = V_{+-}$. We observed previously that the outcome $o$ of deterministic MBQC satisfies $o \cdot g = ++$, where the simulated unitary is $U_{\rm total} = V_g$. Therefore, Alice reports MBQC flux $o^{\rm Alice} = ++$ whereas Bob reports MBQC flux $o^{\rm Bob} = +-$. \bartek{Alice and Bob declare different MBQC fluxes} even though they are describing the same physical experiment. 

\bartek{This example illustrates that MBQC fluxes form a computation-dependent basis for flux space. In particular, MBQC flux is always defined in the context of a specific computation. In contrast, the basis of physical fluxes defined in equation~(\ref{genflux}) does not suffer from a teleological ambivalence; it is computation-independent. The benefit of the MBQC flux basis is that it is directly labeled by computational outputs $o$. Meanwhile, physical fluxes characterize the resource state, independently of the computation for which it may be used. States of definite MBQC flux and physical flux make up distinct bases for the same Hilbert space, the former computation-dependent and the latter computation-independent.} With a fixed choice of the simulated $U_{\rm total}$, changing basis from physical fluxes to MBQC fluxes is described by equation~(\ref{genfluxsimple}). \bartek{We explain in Section~\ref{sec:quantumgauge} that MBQC works} by projecting states of definite physical flux (such as the cluster state) onto states of definite MBQC flux. 

\paragraph{Analogy}
It is useful to compare the definitions of flux (both physical and MBQC) with more familiar gauge theories such as electromagnetism. In $U(1)$ gauge theory, the Aharonov-Bohm effect relates the flux of a magnetic field through a loop to the additional phase incurred by a charged particle that circles that loop:
\begin{equation}
{\rm flux} \,\,\,\propto\,\,\, \oint D_\mu dx^\mu = \oint A_\mu dx^\mu = \rm{phase}
\label{u1wilson}
\end{equation}
Written in this form, the flux \bartek{is measured by the logarithm of a Wilson loop.} The object $D_\mu = \partial_\mu + A_\mu$ is the covariant derivative, which generates parallel transport in the underlying $U(1)$ fiber bundle. In more technical language, $D_\mu$ is the Ehrenfest connection---the generator of parallel transport acting in the total space $\mathcal{E}$. It is the action of $D_\mu$, which is pictured in red in Figure~\ref{holonomy}. 

The object $A_\mu$ is what physicists call the gauge potential. As emphasized before, it is defined relative to a section $f$, which is a map from the base to the fibers. The equality of the two integrals in (\ref{u1wilson}) follows from the assumed single-valuedness of $f$. The two integrands differ by pure coordinate transport $\partial_\mu = D_\mu - A_\mu$ in the base, which lifts in total space to following the section $f$. When the section is single-valued, a closed loop in the base lifts to a closed loop in total space and explicitly including $\partial_\mu$ in (\ref{u1wilson}) becomes immaterial. 

The situation is different when we contemplate a choice of `section,' which is not single-valued. In that circumstance, a closed loop in the base lifts to a non-trivial transformation on the fiber
\begin{equation}
U_{\rm section}\!: f_{\rm initial} \to f_{\rm final} 
\label{nonsinglesec}
\end{equation}
and equation~(\ref{u1wilson}) becomes modified:
\begin{equation}
{\rm flux} \,\,\,\propto\,\,\, \exp \oint D_\mu dx^\mu = \exp \oint A_\mu dx^\mu + \log U_{\rm section}
\label{u1wilson2}
\end{equation}

This relation is directly analogous to equation~(\ref{genflux}), which relates the physical and MBQC flux. Parallel transport defined in (\ref{AF}) plays the role of the covariant derivative $D_\mu$ in continuum gauge theory. The role of the gauge potential\footnote{When the gauge group is continuous---like $U(1)$---the gauge potential is valued in the Lie algebra and its exponentials are group elements. When the gauge group is discrete, we work directly with group elements. Therefore, it is more accurate to say that measurement outcomes $g_j$ are analogous to path-ordered exponentials of $A_\mu dx^\mu$, i.e. Wilson lines of the continuum theory.} is played by the measurement outcomes $g_j$ (in the gauge bundle $\mathcal{G}$) or by their representations matrices $V_{g_j}$ (in the associated vector bundle). The physical flux on the left hand side of (\ref{genflux}) and the MBQC flux $o$ are off by the simulated $U_{\rm total}$. By comparing~(\ref{u1wilson2}) with (\ref{genflux}), we see that $U_{\rm total}$ is analogous to $U_{\rm section}$ in electromagnetism. \bartek{The analogy is subtle because $U_{\rm total}$ in MBQC is actually a superposition of $\mathbb{Z}_2 \times \mathbb{Z}_2$ bundle sections.}

\bartek{To clarify the analogy, we add a comment on the relation between $A_s$ in (\ref{AF}) and the covariant derivative $D_\mu$. In Definition~\ref{GTdef}, condition~(i) we demanded that gauge transformations keep $A_s$ invariant whereas the covariant derivative is, by definition, covariant. At this level, the analogy between $A_s$ and $D_\mu$ is imperfect. To make the analogy fully consistent, we could lift gauge transformations (\ref{gtf}) to act in virtual space as
\begin{equation}\label{sgt2}
\begin{array}{rclcrcl}
U_{f_i} & \mapsto & U_{\tilde{f}_i} =V_h U_{f_i},&& V_{g_i} &\mapsto& V_{\tilde{g}_i} = V_{g_i},\\
U_{f_{i+1}} & \mapsto & U_{\tilde{f}_{i+1}} =U_{f_{i+1}} V_h^\dagger,&&  V_{g_{i+1}} &\mapsto& V_{\tilde{g}_{i+1}} = V_{g_{i+1}}
\end{array}
\end{equation}
instead of equations~(\ref{sgt}). This would affect (\ref{agf1}-\ref{agf2}) and make $A_s = V_g U_f$ covariant, but it would maintain the invariance of the product $A_{g_{i+1}, f_{i+1}} A_{g_i, f_i}$ (up to a phase). Note that this re-identification has no impact in the Hilbert space of physical spins. It only makes a difference in the unobservable virtual space.} 

\bartek{The reason why in conventional gauge theories one does not define an `invariant derivative' (like $A_s$ under gauge transformations~(\ref{sgt})) is that the gauge field typically couples to a physically observable matter field, which also transforms under gauge \bart{transformations}. In MBQC, the `matter field' is purely virtual so we might as well absorb its would-be transformation law in the definition of parallel transport. This is what is done in equations~(\ref{sgt}), in contrast to (\ref{sgt2}). Because of this subtlety, we prefer to use the term `parallel transport' rather than `covariant derivative.'}

\subsection{Elements of computation}
\label{ElemC}

\subsubsection{Output}
\label{Output}
We have the following result:
\begin{Theorem}\label{Holon}
For MBQC on 1D cluster states, in all gauges reachable from $\textbf{f}=0$ by a gauge transformation, the MBQC output $o$ equals the holonomy $g$ of the MBQC gauge potential $\textbf{g}$.  
\end{Theorem}

\noindent
{\em{Proof of Theorem~\ref{Holon}}.} The classical side processing relation~(\ref{cprGTb}) provides $o=g$, by Theorem~\ref{gtT} valid for all gauges $\textbf{f}$ equivalent to 0 up to gauge transformation. Then, $g$ has been shown to be the holonomy of the gauge potential $\textbf{g}$ in Section~\ref{PT}. $\Box$
\medskip

The classical side processing relations are, from the phenomenological perspective, an essential ingredient of MBQCs. Even the simplest MBQC---the three-qubit GHZ-MBQC of Anders and Browne \cite{AB}, repurposing Mermin's star \cite{Merm}---has them. Their presence is a direct consequence of the fundamental randomness in quantum measurement. In MBQC, such randomness must be prevented from affecting the logical processing, and the classical side processing relations ensure that.  Theorem~\ref{Holon} places the classical side processing relation Eq.~(\ref{cprGTb}) in the general theoretical framework of gauge theory. 

\subsubsection{\bartek{Adaptation of} measurement basis from gauge fixing}
\label{TempO}
In Section~\ref{GauP}, we explained that changing the reference $\textbf{f}$---a choice of one intended gate $U_i$ from the set $[U_i]$ at every $i$---does not affect an MBQC computation. For this reason, we identified the reference $\textbf{f}$ with the concept of a section or---in common physics parlance---with the choice of gauge. The reference $\textbf{f}$ is physically undetectable.

As is common in gauge theory, we can exploit the flexibility in choosing $\textbf{f}$ for our convenience by imposing suitable gauge-fixing conditions. A particularly convenient gauge-fixing condition reads:
\begin{equation}
    V_{g_i} = 1 \qquad \textrm{for all}~~i<N
    \label{mbqcbasisgauge}
\end{equation}
Note that the condition is imposed on all blocks except the final one. We cannot impose $V_g = 1$ on \textit{all} blocks because, by equation~(\ref{fluxwords}), this would incorrectly reset the physical flux of the resource state.\footnote{In the continuum, this corresponds to the fact that the gauge potential can always be locally set to zero, but a nonvanishing flux represents an obstruction to achieving such a gauge globally.}
Gauge condition~(\ref{mbqcbasisgauge}) has no independent meaning before a run of MBQC is executed. The choice of reference that is consistent with~(\ref{mbqcbasisgauge}) depends on the registered measurement outcomes. More explicitly, up to (but excluding) the final block, the condition demands that the section value $f_i$ on the $i^{\rm th}$ block be simply equal to the outcome of the $i^{\rm th}$ measurement. Equating the reference with the measurement outcome locally obviates the need for a byproduct, thereby setting $V_{g_i}=1$.

On the final block, the reference can be chosen so that: 
\begin{equation}
    \prod_i U_{f_i} = U_{\rm total}
    \label{lastgaugefix}
\end{equation}
We have seen in equation~(\ref{burelation}) that a complete run of MBQC induces a transformation in virtual space, which equals $B U_{\rm total}$, where $B$ is the byproduct operator. A gauge fixing choice, which best accords with the MBQC interpretation, picks a section which---if measured---ends up simulating $U_{\rm total}$ with no need for further corrections. 

\bartek{If a given section happens to satisfy gauge condition (\ref{mbqcbasisgauge}), it can be described as an ideal MBQC run. By construction, if all MBQC measurements project onto states in the section, no byproduct operators will appear and no basis adaptation will be necessary. If other measurement outcomes do occur, the MBQC technician will be required to adapt the measurement basis; it is also the circumstance when gauge condition~(\ref{mbqcbasisgauge}) dictates a change of section. The two concepts are therefore synonymous: to adapt the MBQC measurement basis is to change the section so as to maintain (\ref{mbqcbasisgauge}). We elaborate on this point below.} 

\paragraph{The adaptive gauge selects the measurement basis} 
An obvious feature of the adaptive gauge (\ref{mbqcbasisgauge}) is that the measurement basis executed in an MBQC run must always include the section value $f_i$, for all $i$. After all, on each block $i < N$, the section $f_i$ gives the measured basis state $|f_i\rangle$. On the final block, the measurement basis certainly contains the state, which induces $U_{\rm total}$ as the transformation on virtual space. That state completes the choice of the adaptive gauge according to equation~(\ref{lastgaugefix}).

The fiber element $f_i$ that is selected by the adaptive gauge-fixing condition gives one state $|f_i\rangle$, which is necessarily part of the measurement basis on block $i$. The full measurement basis over block $i$ is the $\mathbb{Z}_2 \times \mathbb{Z}_2$ multiplet, which is generated from $|f_i\rangle$ by the projective representation $l(g)$. We saw this measurement basis multiplet explicitly in equation~(\ref{gagen}). The section $|f_i\rangle$, from which the multiplet is generated, is the reference state~(\ref{reffi}).

\bartek{We reiterate that when actual measurement outcomes agree with the reference section, no MBQC adaptation is necessary. On the other hand, if actual measurement outcomes disagree with the section, one must adapt the MBQC basis and the gauge choice dictated by (\ref{mbqcbasisgauge}-\ref{lastgaugefix}) changes. We conclude that imposing (\ref{mbqcbasisgauge}-\ref{lastgaugefix}) as a gauge choice and updating the MBQC measurement basis are one and the same thing.}

We must emphasize that identifying a section $f_i$ with the measurement record on $i < N$---and with the adaptive measurement basis everywhere---is a gauge choice, not an identity. Other gauge choices are viable, and indeed necessary, for defining MBQC. When planning an MBQC calculation, a quantum programmer must select a set of reference angles $\alpha_i$, which will be subject to later adaptation. At this initial planning stage (sometimes called pre-compiling), the programmer does not yet know the outcomes of future measurements and so cannot gauge-fix according to (\ref{mbqcbasisgauge}). Pre-compiling crucially requires gauges other than~(\ref{mbqcbasisgauge}). 

\paragraph{MBQC measurement gauge and temporal order}
The relation between the MBQC measurement basis and the reference in the adaptive gauge---the fact that the former always contains the latter---gives an explanation for the temporal order in MBQC. Condition $V_g=1$ (or any other gauge condition) can only be imposed by acting with $r(g) \otimes l(g)$ because those are the generators of gauge transformations \eqref{gtf}. The action of $l(g)$ fixes $V_g=1$, but the associated action of $r(g)$ always affects the section on the subsequent site. Since equation~(\ref{mbqcbasisgauge}) identifies the section with the measurement basis, this is synonymous with the adaptation of the MBQC measurement basis.

Of course, condition~(\ref{mbqcbasisgauge}) can only be applied in a temporally ordered fashion. This is evident from the mechanics of MBQC, but the gauge theory language reveals a deeper explanation. The condition $V_{g_i} = 1$ sets Wilson lines $V_{g_{\prec j}} \equiv 1$ for all $1 \leq j \leq N$. (This includes the gauge-fixing condition~(\ref{lastgaugefix}) because $V_{g_{\prec N}} = B$.) Those Wilson lines are path-ordered products of individual $V_{g_j}$s. Setting $V_{g_{\prec j}} \equiv 1$ can only be done in a time-ordered fashion: first $V_{g_{\prec 1}} \equiv V_{g_1}$, then $V_{g_{\prec 2}} \equiv V_{g_2} V_{g_1}$, etc.

\section{Phenomenological ramifications of the gauge principle}
\label{equiv}

\subsection{Equivalent measurement records}
\label{sec:emr}
As reviewed in Section~\ref{sec:review}, there are classes of distinct measurement records that yield the same computational output. This is reflected in the classical side processing relations. Here we show how the MBQC gauge transformations, together with the MPS tensor symmetry $l(\mathbb{Z}_2\times \mathbb{Z}_2)$, imply this property. Specifically, we show that the  change in the measurement record $\textbf{g}=(g_1,g_2,..,g_N)$ given by 
\begin{equation}\label{stild}
\textbf{g} \to (g_1,..g_{i-1},h^{-1}g_i,h g_{i+1},g_{i+2},..,g_N)
\end{equation}
leaves the circuit simulation invariant for all $h \in \mathbb{Z}_2\times \mathbb{Z}_2$.  These transformations, with $h$ acting on all adjacent block pairs except $(N,1)$, generate a class of equivalent outcomes.

To show this, we subject the product $A_{\textbf{g},\textbf{f}}$ to the MBQC gauge transformation~(\ref{sgf}) and (\ref{sgt}). Only the product $A_{g_{i+1},f_{i+1}}A_{g_i,f_i}$ is potentially non-trivially affected, and so we focus on that part:
$$
\begin{array}{rcl}
A_{g_{i+1},f_{i+1}}A_{g_i,f_i} &=& \left(V_{g_{\prec i+1}} \left( V_{g_{i+1}} U_{f_{i+1}}\right) V_{g_{\prec i+1}}^\dagger \right) \left(V_{g_{\prec i}} \left( V_{g_i} U_{f_i}\right) V_{g_{\prec i}}^\dagger \right)\\
& \mapsto& \left(V_{\tilde{g}_{\prec i+1}} \left( V_{\tilde{g}_{i+1}} U_{\tilde{f}_{i+1}}\right) V_{\tilde{g}_{\prec i+1}}^\dagger \right) \left(V_{\tilde{g}_{\prec i}} \left( V_{\tilde{g}_i} U_{\tilde{f}_i}\right) V_{\tilde{g}_{\prec i}}^\dagger \right)\\
&\propto & \left(V_{\tilde{g}_{\prec i+1}} \left( V_{g_{i+1}} V_h U_{f_{i+1}} V_h^\dagger \right) V_{\tilde{g}_{\prec i+1}}^\dagger \right) \left(V_{\tilde{g}_{\prec i}} \left( V_{g_i} U_{f_i}\right) V_{\tilde{g}_{\prec i}}^\dagger \right)\\
&\propto & \left(V_{\tilde{g}_{\prec i+1}}\left( V_h \left( V_{g_{i+1}}  U_{f_{i+1}}  \right) \right) V_{\tilde{g}_{\prec i+1}}^\dagger \right) \left(V_{\tilde{g}_{\prec i}} \left( V_h^\dagger \left( V_{g_i} U_{f_i}\right)\right) V_{\tilde{g}_{\prec i}}^\dagger \right)\\
&= & \left(V_{\tilde{g}_{\prec i+1}}  V_{h g_{i+1}}U_{f_{i+1}}  V_{\tilde{g}_{\prec i+1}}^\dagger \right) \left(V_{\tilde{g}_{\prec i}}  V_{h^{-1} g_i}U_{f_{i}} V_{\tilde{g}_{\prec i}}^\dagger \right)\\
&=&  A_{h g_{i+1},f_{i+1}} A_{h^{-1} g_i,f_{i}}
\end{array}
$$
In the second line we applied the gauge transformation $h^{(i)}$ according to~eq.~(\ref{sgt}), in the third line we unravelled it, in the fourth line we reordered operators that commute up to phase.  In the fifth line and the sixth line, we used the left symmetry  $l(h)$ and noted that adaptation in the $i+1$ block given by the conjugation of $V_{\tilde{g}_{\prec i}}$ is consistent with the change in measurement outcome $g_{i} \to h^{-1} g_{i}$ on the previous block. 

Thus we have shown that measurement records related by
\begin{align}\label{Emeas}
    g_{i+1} &\to h g_{i+1} ,\quad    g_{i}\to h^{-1}g_{i}   \nn
    f_{i+1} &\to f_{i+1}  \qquad f_{i} \to f_{i}
 \end{align}
 are equivalent with respect to the quantum computation.  We emphasize that unlike the gauge transformations, in \eqref{Emeas} the group labels change while the references are fixed. Consequently, the gauge invariant outcomes $  s_{i}=g_{i}f_{i} $ transform as
\begin{align}
    s_{i} \to h^{-1} s_{i},\qquad s_{i+1} \to h s_{i+1},   
\end{align}
whereas all other measurement outcomes remain unchanged.

\subsection{MBQC holonomy and the virtual circuit simulation }
For simplicity, here and in the next subsection we consider the special case of deterministic MBQC, i.e., where one of the four possible output values $o$ occurs with unit probability. This means that
$$
U_\textbf{f} \in [I].
$$ 
Note that this does not require that all local measurement bases are the trivial $X$-basis. The Pauli unitary $U_\textbf{f}$ may be accumulated in small (non-Clifford) chunks.

The cluster state resolved in the measurement basis is:
$$
|1\rangle = \mathcal{N} \sum_{\textbf{g}} \text{Tr}(A_{\textbf{g},\textbf{f}})\,
|\textbf{g},\textbf{f}\rangle.
$$
With eq.~(\ref{gf}) and the case assumption $U_\textbf{f} \in [I]$, it holds that $A_{\textbf{g},\textbf{f}} \in [I]$, for all (potential) measurement records $\textbf{g}$. Therefore, for all measurement records $\textbf{g}$ that can actually occur with non-zero probability, it follows that $A_{\textbf{g},\textbf{f}}=I$. Using~(\ref{gf}) again, we find that 
\begin{align}\label{UVo}
    U_\textbf{f} = V_o^{-1} \propto V_o,
\end{align}
for all measurement records that occur with non-zero probability. Thus, the holonomy $U_\textbf{f}$ of the virtual quantum register is cancelled by the holonomy $o=g$ of the gauge potential $\textbf{g}$. This goes to the essence of the MPS circuit model interpretation \cite{Eis2} of MBQC: the circuit model quantum register, living on the virtual (or correlation) space, is never directly probed. It interacts with the physical spins, and those are being measured. The measurement record gives rise to the holonomy $g$, which equals the MBQC output. Thus, the virtual quantum register is measured indirectly, relying on the above cancellation of holonomies.

Non-deterministic simulations require a more detailed analysis, which involves quantum superpositions of fluxes. That case will be covered in Section~\ref{sec:quantumgauge}.

\subsection{Computational output and homotopy}
Here we introduce the notion of MBQC order parameter, and discuss how the non-Abellianness of the first homotopy group of the order parameter space relates to MBQC temporal order.

We observed in~(\ref{class}) that in the MBQC implementation of every gate $U_i$ we can only aim for the equivalence class $[U_i]$. This suggests that it is meaningful to consider $[U_{\prec t}]$ as an order parameter, evolving in time $t$. The possible order parameters form the manifold ${\cal{M}}$.

With an eye on homotopy, we want the time parameter to be continuous. This is achieved as follows: for an integer time $i$, $i\leq t \leq i+1$, the measurement angles $\alpha_{i,1}$ and $\alpha_{i,2}$ are ramped up continuously from $\alpha_{i,1}=\alpha_{i,2}=0$ to their intended values as $t$ increases from $i$ to $i+1$. In this way, we are mapping the interval $[0,N]\ni t$ to a trajectory in the parameter space ${\cal{M}}$. In our MBQC on a cluster ring, this parameter space is
\begin{equation}\label{OPspac}
{\cal{M}} = SU(2)/{\cal{Q}}.
\end{equation}
Herein ${\cal{Q}}=\{\pm I, \pm iX, \pm iY, \pm iZ\}$ is the quaternion group, which is the lift of the group $\mathbb{Z}_2\times \mathbb{Z}_2$ into $SU(2)$.

For deterministic MBQCs, $[U_{\prec t}]$ is the same at $t=0$ and $t=2$. We may therefore identify the two points, and obtain a mapping from the circle into ${\cal{M}}$. These maps, with base point at $t=0$, are classified by the first homotopy group $\pi_1({\cal{M}})$. They are also classified by the total (Pauli) unitary accumulated in $U_{\prec N}\in SU(2)$ at time $t=N$, and it is the same classification.

For our order parameter space of eq.~(\ref{OPspac}), the first homotopy group is:
\begin{equation}\label{pi1}
\pi_1(SU(2)/{\cal{Q}}) = {\cal{Q}}.
\end{equation}
The total Pauli operator accumulated during the evolution from continuous time $t=0$ to $t=N$ is not visible in ${\cal{M}}$ at any given instant of time, because it is modded out in ${\cal{M}}$. But it is captured in the trajectory in ${\cal{M}}$ as a whole, specifically as the homotopy class of the trajectory. 

We learn: In case of deterministic MBQC, the total unitary accumulated is an element of the homotopy group $\pi_1({\cal{M}})$. This group element contains the computational output $o$, and in addition a global phase $\pm 1$.

The phase $\pm 1$ shows up in the classification but is not directly of physical interest. So one may wonder whether it is a nuisance or a feature. There is indeed a role that the phases play---they are a signature of the non-Abelianness of the homotopy group ${\cal{Q}}$. Is the non-Abelianness of ${\cal{Q}}$ of significance for MBQC? 

The elements of ${\cal{Q}}$ are the byproduct operators, which are also the generators of the simulated unitary evolution. We recall that the forward-propagation of byproduct operators is responsible for temporal order in MBQC. And this is precisely where the commutation relations of the byproduct operators come into play. There is non-trivial temporal order if and only if the byproduct operators do not commute.

\begin{figure}
\begin{center}
\includegraphics[width=9cm]{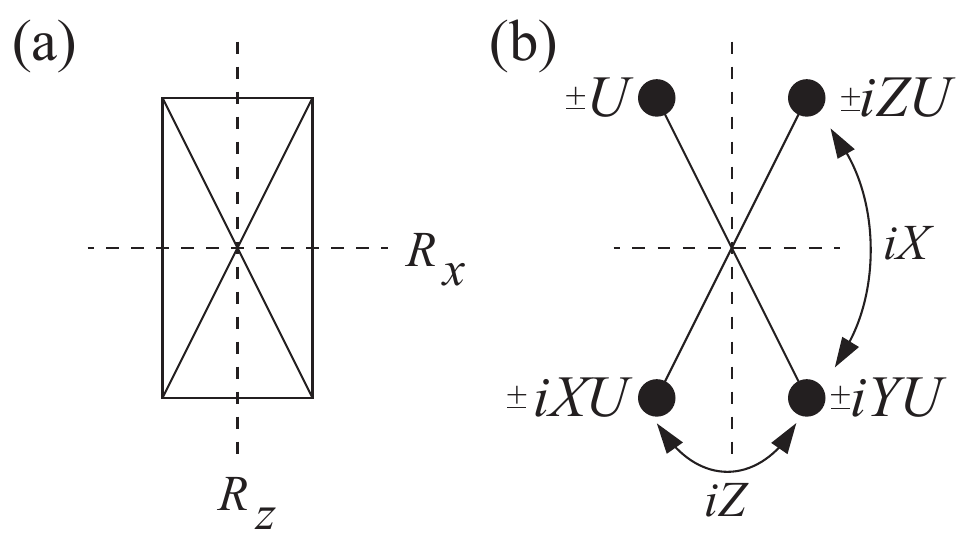}
\caption{\label{order}The order parameter spaces for biaxial nematics and MBQC on a cluster ring are the same, ${\cal{M}}=SO(3)/(\mathbb{Z}_2\times \mathbb{Z}_2)=SU(2)/{\cal{Q}}$. (a) Order parameter of biaxial nematics. The symmetry is realized by reflections about the horizontal and the vertical axis. (b) Order parameter of MBQC on a cluster ring. The symmetry is implemented by left-multiplication with elements of ${\cal{Q}}$.  }
\end{center}
\end{figure}

\paragraph{Analogy.} A system from classical physics with the same order parameter space ${\cal{M}} = SO(3)/(\mathbb{Z}_2\times \mathbb{Z}_2)=SU(2)/{\cal{Q}}$ are the biaxial nematics \cite{Merm2}. There, too, the non-Abelianness of $\pi_1({\cal{M}})$ matters. Namely, it gives rise to non-trivial interaction among linear defects in 3D crystals of such materials. See Fig.~\ref{order} for a comparison of the two realizations of the order parameter space.

\subsection{A quantum MBQC gauge theory}
\label{sec:quantumgauge}

A classical gauge theory is described by a principal fiber bundle on which every path $\gamma$ in the base space picks out a definite holonomy.  In contrast, a quantum gauge theory acommodates (and requires) superposition of states with different holonomies.  For example, such superpositions are a basic feature of quantum electrodynamics, in which Wilson loops are fluctuating observables.

Likewise, in  MBQC, holonomies are superposed in order to  simulate a general unitary  $U \in SU(2)$ in a probabilistic fashion.  More precisely, given an expansion 
\begin{align}\label{Uex}
    U = \sum_{o} c_{o} V_{o},\qquad o \in \mathbb{Z}_{2} \otimes \mathbb{Z}_{2}
\end{align}
the cluster state can be expanded in terms of a set of $U$-dependent, entangled states $\ket{U^{\dagger}V_{o}}$ with holonomy $o$: 
\begin{align}\label{1o}
    \ket{1} = \sum_{o} c_{o} \ket{U^{\dagger}V_{o}}
\end{align}
Equations \eqref{Uex} and \eqref{1o}  generalize the relation \eqref{UVo} between $o$ and the $U$  to allow for quantum superpositions.  The probabilities $|c_{o}|^{2}$ provide a sort of state-operator correspondence, in which logical states $\ket{U^{\dagger}V_{o}}$ in the physical Hilbert space are mapped to operators $U$ in the virtual space.\footnote{Actually MBQC only determines the modulus of $c_{o}$ and not the relative phases. This feature is explained below in terms of the decomposition of the Hilbert space into superselection sectors.}  In this section, we explain the precise nature of this correspondence.  In particular, we will show how the logical states $\ket{U^{\dagger}V_{o}}$ arise from a decomposition of the physical Hilbert space into flux sectors, and how MBQC uses this decomposition to implement quantum computations.

\subsubsection{Superselection (flux) sectors} 
\label{sec:defflux} 

\paragraph{Superselection sectors in the $X$-basis}
 In Section~\ref{sec:review} we showed that the holonomy $o$ is invariant under changes of the measurement record generated by stabilizers.    These transformations permute the computationally equivalent measurement outcomes described in Section~\ref{sec:emr}.  They determine a basis, which forms an orbit under the action of the stabilizers
\begin{align}
    \mathcal{L} =\{ K_{i}, i=2,\cdots N-1\}.
\end{align}
We exclude the stabilizers $K_{1}$ and $K_{N}$ because they are not needed to span the measurement outcomes, and are also inconsistent with the temporal order on the circle.\footnote{ The temporal order in MBQC selects out a `base point' on the circle, which separates the past from the future.  Indeed, measurement outcomes at the $j=N$ block should not affect measurement angles at the $j=1$ site. The stabilizers $K_{1},K_{N} $ violate this constraint.}
 
We define a flux sector to be the span of such an orbit.  Each flux sector is labelled by the holonomy $o$, so that the total Hilbert space  $\mathcal{H}_{S^{1}}$ on a circle  decomposes into a direct sum
\begin{align}
\label{HS1}
\mathcal{H}_{S^1} =\oplus_{o} \mathcal{H}_{o}
\end{align}  
As an example, for measurements along $X$, the basis for $\mathcal{H}_{o}$ comprises $X$-eigenstates:
\begin{align}
    |g_{N}\rangle \otimes \ldots \otimes |g_1\rangle , \qquad
    o = \prod_{i}g_{i}
\end{align} We first encountered these sectors in Section~\ref{cluster}, in the paragraph `A glimpse of a global structure.'   

\paragraph{Twisted states} 
We continue to concentrate on $X$-measurements, postponing the general case till the next paragraph.
In each sector labeled by $o \in \mathbb{Z}_2 \times \mathbb{Z}_2$, there is a distinguished representative, which can be expressed as an entangled matrix product state:
\begin{equation}
\label{Vh}
|V_o\rangle := 
\mathcal{N}
\sum_{g_j \in \mathbb{Z}_2 \times \mathbb{Z}_2} 
{\rm Tr} \big(V_o A_{g_{N}} \ldots  A_{g_{2}}A_{g_{1}}\big) 
|g_{N}\rangle \ldots |g_{2}\rangle |g_{1}\rangle.
\end{equation}
Notice that the cluster state, as defined in~(\ref{mpsring}), is (\ref{Vh}) with $V_{o}=1$. This is the precise reason for denoting the cluster state as $\cluster$. 

Equation~(\ref{Vh}) encompasses three other matrix product states---$|X\rangle$, $|Z\rangle$, and $|XZ\rangle$---which are canonical representatives of the three other superselection  sectors. Their wavefunctions overlap only with those states $|g_{N}\rangle \ldots |g_{2}\rangle |g_{1}\rangle$ whose labels satisfy
\begin{equation}
g_N \ldots g_2\, g_1 = o.
\end{equation}
Notice that any of these states can be used as a resource for MBQC.   In particular, measurements of $|V_o\rangle$ along $X$ give the deterministic computational output $o$. 

Including a nontrivial $V_o$ in the wavefunction~(\ref{Vh}) is an insertion of physical $\mathbb{Z}_2 \times \mathbb{Z}_2$ flux. We will call states $|V_o\rangle$ `twisted states' because a flux insertion corresponds to imposing a `twisted boundary condition' on a matter particle living in virtual space. In a continuum gauge theory, such twisted boundary conditions are implemented by coupling a matter field $\Phi(x)$ to the gauge field via 
$
\Phi(x) \to e^{i \int^{x} A_\mu dx^\mu }\Phi(x)  
$
and demanding:
\begin{align}
\Phi(x + 2 \pi ) &= e^{i \oint A}\Phi(x)
\end{align} 

\paragraph{Flux sectors for a general measurement basis}
We can apply the same logic to decompose $\mathcal{H}_{S^{1}}$ into sectors $\mathcal{H}_{o}(U)$ defined with respect to a general, $U$-dependent measurement basis:
\begin{align}
\bigotimes_{j=1}^{2N} |(-)^{s_j}_{\big((-)^{q_{j}} \alpha_{j}\big)}\rangle
= 
\bigotimes_{i=1}^N |g_i\rangle
\label{mbqchistory}
\end{align} 
A projection of the cluster state onto a product state in~(\ref{mbqchistory}) represents one measurement history of an MBQC computation, which is designed to simulate a unitary $U$. These histories, too, fall into four orbits under $\mathcal{L}$. Each orbit is labeled by the holonomy $o=\prod_{i}g_{i}$,  where the group labels $g_i \in \mathbb{Z}_2 \times \mathbb{Z}_2$ are relative to a section in the adapted basis.

Following~(\ref{Vh}), we define twisted matrix product states, which have definite holonomy as measured by a run of MBQC:
\begin{equation}
\label{Vhp}
|U^{\dagger}V_o\rangle := 
\mathcal{N}
\sum_{g_j \in \mathbb{Z}_2 \times \mathbb{Z}_2} 
{\rm Tr} \big(U^{\dagger} V_o A_{g_{N}}\ldots  A_{g_{2}}A_{g_{1}} ) 
|g_{N}\rangle \ldots |g_{2}\rangle |g_{1}\rangle 
\end{equation}
Note that this notation is consistent with the definition in \eqref{Vh} because the combination 
\begin{align}
    \sum_{g_j \in \mathbb{Z}_2 \times \mathbb{Z}_2}
\big( A_{g_{N}} \ldots  A_{g_{2}}A_{g_{1}} \big) |g_N\rangle \ldots |g_2\rangle |g_1\rangle
\end{align}
 is basis-independent. So equation \eqref{Vhp} can be read with $\ket{g} $ defined relative to any measurement basis.  In particular, for measurements in the $X$-basis, the twisted states \eqref{Vhp} reduce to equation~(\ref{Vh}) with $V_{o}$ replaced by $U^{\dagger} V_{o} $.

The twisted states \eqref{Vhp} can be characterized by the fact that running an MBQC simulation of $U$ on them would return a deterministic outcome.  Indeed, using \eqref{burelation}, we can express the wavefunction of these twisted states in the adapted basis as:
\begin{align} 
{\rm Tr} \big(U^\dagger V_o A_{g_{N}} \ldots  A_{g_{2}}A_{g_{1}} \big) =
  {\rm Tr} \big(V_{o} B_{{g_N} \ldots g_2 g_1}\big) \propto \delta_{o, g_N \ldots g_2 g_1}
\end{align}
Thus, state $|U^{\dagger}V_o\rangle$ has definite MBQC holonomy $o=g_N \ldots g_2 g_1$. 

\subsubsection{A logical Hilbert space of fluxes}
The resource states for MBQC live in the $2^{2N}$-dimensional Hilbert space $\mathcal{H}_{S^{1}}$ of physical qubits.  On the other hand, the twisted states span a four-dimensional Hilbert space
\begin{align}
    \mathcal{H}_{\text{flux}}=\text{span} \{ \ket{V_o} \}
\end{align} 
whose kets represent flux sectors of the gauge theory.  \bartek{Schematically, we have $\mathcal{H}_{\text{flux}} = \mathcal{H}_{S^1} / \textrm{(stabilizers)}$, as is usual in the quotient construction of gauge theory Hilbert spaces.} MBQC uses $\mathcal{H}_{\text{flux}}$ as a logical Hilbert space to perform quantum computations.   In particular, MBQC makes use of the freedom to rotate the basis of twisted states  \eqref{Vh} into \eqref{Vhp}.  We can make the unitary rotation explicit by reading both \eqref{Vh} and \eqref{Vhp} in the $X$-basis, and then expanding:
\begin{equation}
U^\dagger  
:= \sum_{k \in \mathbb{Z}_2 \times \mathbb{Z}_2} \tilde{c}_{k} V_{k}
\end{equation}
The linearity of the matrix product states then immediately implies:
\begin{equation}
|U^{\dagger} V_o\rangle = 
\mathcal{N}\!\!\!\!
\sum_{k \in \mathbb{Z}_2 \times \mathbb{Z}_2} \tilde{c}_k
\sum_{g_j \in \mathbb{Z}_2 \times \mathbb{Z}_2} 
{\rm Tr} \big(V_k V_o A_{g_{N}} \ldots  A_{g_{2}}A_{g_{1}} \big) 
|g_{N}\rangle \ldots |g_{2}\rangle |g_{1}\rangle
= \mathcal{N}
\sum_k \tilde{c}_k\, b(k,o)\, |V_{ko}\rangle,
\label{changebase}
\end{equation}
where $b(k,o)$ is the projective phase defined by $V_k V_o = b(k,o) V_{ko}$. 

We thus have two flux bases for $\mathcal{H}_{\text{flux}}$. On the one hand, we have a canonical set $|V_o\rangle$ to which the cluster state $\cluster$ belongs; these are  resource states that an MBQC technician will hold in her lab. On the other hand, we have $\ket{U^{\dagger} V_{o}}$, defined so that a $U$-simulating MBQC protocol conducted on the initial state $\ket{U^{\dagger} V_{o}}$ would return a definite holonomy (deterministic computational output.) While the latter are not resource states, they are nevertheless essential in MBQC.  

To see this, let us revisit the MBQC protocol on a ring from the perspective of  $\mathcal{H}_{\text{flux}}$.   We can expand the cluster state $\ket{1}$ in terms of $\ket{U^{\dagger} V_{o}}$ by applying the inverse of \eqref{changebase}, or equivalently, by expanding $U= \sum_{o} c_{o} V_{o}$ in equation \eqref{1}.  This gives:
\begin{align}\label{1'}
    \mathcal{N}^{-1}
    \ket{1} &= \sum_{g} {\rm Tr} (U B_{g_{N} \ldots g_{2} g_{1} } ) |g_{N}\rangle \ldots |g_{2}\rangle |g_{1}\rangle
\nn    &= \sum_{o}c_{o} \big( \sum_{g} {\rm Tr} (V_{o} B_{g_{N} \ldots g_{2} g_{1} } ) |g_{N}\rangle \ldots |g_{2}\rangle |g_{1}\rangle\big)\nn
    &=\sum_{o}c_{o}  \ket{U^{\dagger} V_{o} }
\end{align}
Recall from Section~\ref{sec:mbqcring} that the raw output of an MBQC calculation on a ring is a probability distribution  $p_o = |c_o|^2$, which characterizes the simulated unitary $U$. Thus, even though MBQC is physically implemented by single qubit measurements,  from a global perspective it is measuring the MBQC flux corresponding to $\ket{U^{\dagger} V_{o}}$.   

\subsubsection{MBQC versus the circuit model}
\label{sec:contrast}
Equation \eqref{1'} presents the cluster state as an element of the four-dimensional flux Hilbert space $\mathcal{H}_{\text{flux}}$.  This perspective provides an important insight into the relation between MBQC and the circuit model.  To illustrate this relation, consider the following hypothetical realization of the circuit model.  Suppose we can create arbitrary superpositions of the twisted states $\ket{ V_{o}}$. We could then effectively apply a unitary rotation on $\mathcal{H}_{\text{flux}} $ such that  
\begin{align}
    \ket{1} \to \ket{U} = \sum_{o} c_{o} \ket{V_{o}}
\end{align}
In the $X$-basis, the resulting resource state is:
\begin{align}
    \label{U}
\ket{U}      = 
\mathcal{N} \sum_{g_{1},\cdots g_{n}} \tr (U A_{g_{N}}\cdots A_{g_{1}} ) \ket{g_{N}} \cdots \ket{g_{1}} 
\end{align}
The form of this MPS wavefunction implies that single qubit measurements of $\ket{U}$ along $X$ will simulate the unitary $U$.  The computational output is
\begin{equation}
p_o= | \langle V_o | U \rangle |^2 = |c_o|^2.
\end{equation}
From the point of view of the flux Hilbert space $\mathcal{H}_{\text{flux}}$, this protocol is a realization of the  circuit model.  

Let us contrast this with MBQC. Here, too, we start with $\cluster$ as the initial state.   However, instead of applying unitary gates to $\cluster$, we measure it in the basis $|U^{\dagger} V_o\rangle$.  Equation (\ref{changebase}) implies these measurements produce the same probability distributions as above:
\begin{equation}
p_o = |\langle U^\dagger V_o| 1\rangle |^2 = |c_o|^2 = | \langle V_o | U \rangle |^2 
\label{UV1}
\end{equation}
In this way, MBQC can be viewed as a `passive' counterpart to an `active' circuit model. Whereas the circuit model implements its programming via the active transformations  $\cluster \to |U\rangle$, MBQC implements its programming via a $U$-dependent rotation of the measurement basis.  In particular,  from the viewpoint of the logical Hilbert space $\mathcal{H}_{\text{flux}}$, MBQC measurements project $\ket{1}$ onto the rotated states $|U^\dagger V_o  \rangle$. Meanwhile, from the persective of the physical Hilbert space $\mathcal{H}_{S_{1}}$, MBQC  projects $\ket{1}$ onto a superselection sector $\mathcal{H}_{o}(U)$ represented by $|U^\dagger V_o  \rangle$.  This superselection sector is defined by the $U$-dependent decomposition
\begin{align} 
\label{HS2}
\mathcal{H}_{S^1} =\oplus_{o} \mathcal{H}_{o}(U),
\end{align}  
which is determined by the change of basis \eqref{changebase} from $\ket{V_{o}}$ to $\ket{U^{\dagger} V_{o} }$.  This rotation is responsible for introducing temporal order into the measurement basis.

\subsection{Extension to fixed point of SPT phases with onsite symmetries}
\label{sec:spt1}
In the preceding sections, we have formulated  all elements of the MBQC gauge theory in terms of the symmetry actions $l(g)$, $r(g)$, and $u(g)$, as well as the equivariant transformation laws of the cluster MPS tensor; see Figure~\ref{equivar}.  These symmetry properties are not unique to the cluster state.  

Indeed, they are shared by all resource states in the SPT phase because \bart{the cluster state is a coarse-grained description of all SPT-ordered states \cite{Schuch}. Its MPS tensor, which satisfies the equivariant transformation laws shown in Figure~\ref{equivar}, is a fixed point of RG flow. What this means in practice is that every state in the SPT phase is related to the fixed point (cluster) state by a finite-depth unitary, which is itself symmetric under the protecting symmetry (here $\mathbb{Z}_2 \times \mathbb{Z}_2$). Therefore, for any state in the phase, we can define analogues of the representations $l(g)$ and $r(g)$ by conjugating equations (\ref{lgprojective}) and (\ref{rgprojective}) by some requisite finite-depth unitary. To apply our discussion to a generic state in the SPT phase, we simply replace $l(g)$ and $r(g)$ with their fine-grained (finite depth circuit-conjugated) analogues.}

\subsection{Extension of the gauge principle to SPT cluster phase}
\label{ESPT}
The 1D cluster state is surrounded by an SPT phase with global symmetry $\mathbb{Z}_2\times \mathbb{Z}_2$ \cite{SPT1,SPT2,SPT3}. It is known that MBQC computational power extends from the 1D cluster state into all of that phase \cite{SPT4}. What about the MBQC gauge principle---does it extend as well? In this regard, we have the following result:

\begin{Theorem}\label{gtTspt}
For any given MBQC on a ground state in the $\mathbb{Z}_2\times \mathbb{Z}_2$ SPT phase surrounding the 1D cluster state, the measurement bases $|g_i,f_i\rangle$ chosen as a function of prior measurement record $g_{\prec i}$, the output $o$ given the measurement record $\textbf{g}$, and the probability distribution of the computational output $o$, $p(o)= |\text{Tr}(U_\textbf{f} V_o)|^2$, are all invariant under the gauge transformations~(\ref{sgt}-\ref{gtg}). 
\end{Theorem}
\noindent
The proof of Theorem~\ref{gtTspt} is the same as of Theorem~\ref{gtT}.\smallskip

{\em{Remark:}} The reason that the proofs of Theorems~\ref{gtT} and \ref{gtTspt} are the same is that both theorems are chiefly about the classical side processing in MBQC (items (i) and (ii)), and this classical side processing extends from the cluster state into the surrounding SPT phase without change \cite{SPT4}. The invariance of $p(o)$ in item (iii) also follows 
as before, but the equality $p(o)=|\text{Tr}(U_\textbf{f} V_o)|^2$ needs to be established for the SPT phase first; this was done in \cite{SPT4}.

There is a lesson to learn from the ease with which the MBQC gauge principle extends from the cluster point into the surrounding phase. The gauge theory of MBQC is, in a certain sense, complementary to `computational phases of quantum matter,' i.e., to viewing MBQC through the lens of symmetry-protected topological order. Computational phases of matter are about quantum states. The MBQC gauge theory is, in turn, about observables.

\section{Conclusion}
The main thesis of this work is that the workings of MBQC can be recast in terms of concepts from gauge theory. The principal identifications are as follows:
\begin{enumerate}
    \item There are many ways to describe the same MBQC computation to a lab technician. They are related by gauge transformations; see Section~\ref{GauP}.
    \item Under a suitable gauge-fixing condition (equations~\ref{mbqcbasisgauge}-\ref{lastgaugefix}), different gauges also correspond to distinct but computationally equivalent measurement records. Imposing this gauge-fixing condition determines the temporal order in the adaptation of measurement angles; see Section~\ref{TempO}.
    \item Computational output is a holonomy of the gauge field or, equivalently, a flux; see equation~(\ref{defmbqcflux}).
\end{enumerate}
We hope that this correspondence helps to bridge a gap between the quantum computing community and high energy and condensed matter theorists, for whom gauge theory is an essential tool. From the condensed matter perspective, a connection between MBQC and gauge theory could have been anticipated: \bartek{many} MBQC resource states have SPT order \cite{SPT0,SPT1,SPT2,SPT3,SPT4,SPT5,SPT5b,SPT6,SPT7}, and SPT order has been characterized in gauge theoretic terms \cite{Levin2012}. In this sense, our work completes the missing link in the triangle: MBQC-SPT-gauge theory.

Several features of the gauge theoretic description of MBQC may seem unfamiliar or even exotic when compared with a standard textbook on gauge theory. The foremost of them is that MBQC encodes the simulated unitary $U$ using linear superpositions of distinct physical fluxes; see equation~(\ref{Vhp}). Such flux superpositions give our gauge theory the flexibility to describe all MBQC computations (in one dimension) in one unified language. Different simulated $U$s lead to distinct notions of MBQC flux, but each of them is unitarily related to physical flux; see equation~(\ref{defmbqcflux}).

One innovation, which made our analysis possible, was to consider as resources cluster states with the topology of a circle. On a technical level, this is what allowed us to define fluxes---that is, holonomies of a fiber bundle. But the novelty of considering circular resource states also brings about a modest practical benefit.
In contrast to MBQC on a cluster with endpoints, MBQC on a ring-like resource extracts two classical bits $o \in \mathbb{Z}_2 \times \mathbb{Z}_2$ of computational output. This agrees with the common lore (superdense coding) that one qubit is worth two classical bits.

Circular resource states are also intriguing from a more conceptual point of view, which concerns the relationship between gauge theory and teleportation. We have motivated our analysis by explaining that MBQC in one dimension is a sequence of gate teleportations. Our identifications between MBQC and gauge theory concepts therefore have direct meaning in quantum teleportation:
\begin{enumerate}
    \item Physical parallel transport~(\ref{AF}) is the gate, which is induced by a projective measurement in quantum gate teleportation.
    \item The intended gate in gate teleportation depends on the intention of the sender and recipient, so it is not gauge invariant. Different intended gates correspond to different gauges---different reference states $|f\rangle$ for the projective measurements.
    \item The classical information sent by Alice to Bob concerns how the measured state $|g\rangle$ differs from $|f\rangle = l(g)^{-1} |g\rangle$. That is, the classical message exchanged in quantum teleportation is an analogue of the gauge potential. 
\end{enumerate}
These identifications extend and complement similar observations about quantum teleportation, which were made in \cite{withlenny}; see also \cite{mielczarek}. Ref.~\cite{withlenny} argued that quantum teleportation with post-selection has the same physical effect as parallel transport of a charged particle in a background gauge field. This is a special case of our identifications, restricted to the circumstance where no corrective operation is necessary. The present work generalizes \cite{withlenny} by removing the post-selection assumption, and by explaining that the corrective operator applied by Bob (the recipient) plays the role of the gauge connection. It also explains how the `entanglement holonomies' defined in \cite{withlenny} can be constructed and measured in a lab. To close an `entanglement holonomy,' the authors of \cite{withlenny} needed a sequence of teleportations, with the initial state of the first teleportation identified with the final state of the last one. MBQC can realize this setup without closing a loop in time and without post-selection---by using a resource state with a ring-like topology. 

\paragraph{Future directions.}
This paper covers MBQC on one-dimensional resource states---settings, which afford simulation of single-qubit gates. For universal quantum computation, one also needs an entangling gate. To simulate it, MBQC requires a two-dimensional resource state.  \bartek{An MBQC run based on a two-dimensional cluster state starts out by projecting out many spins via $Z$-measurements, which reveals a piecewise one-dimensional structure that is connected at triple junction points. Upgrading our discussion to this more elaborate context---which appears to involve Wilson lines connected with three-way junctions}---is the obvious next step of our program.

Another specialization adopted in this paper is that we mostly focused on MBQC on circular resource states. They are best suited for the gauge-theoretic treatment because they naturally support holonomies / fluxes, i.e. Wilson loops. But in its conventional form, MBQC assumes a resource state with the topology of a line segment. Na{\"\i}vely, designing a gauge-invariant computational output on a line segment is challenging because the natural objects---Wilson lines---are not gauge-invariant. The solution involves St{\"u}ckelberg fields / edge modes of the gauge theory \cite{Buividovich:2008gq,Donnelly:2014gva,Donnelly:2016auv}, whose transformations cancel out the gauge dependence of a Wilson line at its endpoints. We postpone a detailed description of conventional MBQC in terms of gauge theory with Stuckelberg fields till a future publication.

A different conceptual perspective, which we mentioned only briefly in the current work, would take the symmetry protected topological (SPT) order of the resource state as an analytic starting point. SPT-ordered phases of matter are in one-to-one correspondence \cite{Shiozaki:2016cim} with equivariant topological field theories (TFT) \cite{Moore:2006dw}. This correspondence suggests elevating the connection between MBQC and gauge theory to one with topological field theory. That, in turn, should bring to focus aspects of MBQC that are more germane to the spirit of a topological field theory, for example scale invariance or an axiomatic definition. An inspection of MBQC from this perspective was initiated in \cite{gabrielnew}.

\section*{Acknowledgements}
GW acknowledges support by Fudan University, the Thousand Talents Program, and Harvard CMSA. RR is supported by NSERC, in part trough the Canada First Research Excellence Fund, Quantum Materials and Future Technologies Program, and by US ARO (W911NF2010013). BC is supported by an NSFC grant number 12042505, a BJNSF grant under the Gao Cengci Rencai Zizhu program, and a Dushi Zhuanxiang Fellowship. We thank the organizers of workshop `Quantum Information and String Theory 2019' held at Yukawa Institute for Theoretical Physics, Kyoto University, where this work was initiated. We also acknowledge the workshop `Symmetry, Phases of Matter and Quantum Computing -- ASQC' (2019) held at Perimeter Institute and BC acknowledges the workshop `Reconstructing the Gravitational Hologram with Quantum Information' (2022) held at the Galileo Galilei Institute in Florence, where this work was partly carried out. 

\appendix

\bibliographystyle{utphys}

\end{document}